# Quantum Theory of Exciton Magnetic Moment: Interaction and Quantum Geometric Effects

**Authors:** Gurjyot Sethi[1,2], Jiawei Ruan[1,2], Fang Zhang[1,2], Weichen Tang[1,2], Chen Hu[1,2], Mit Naik[1,2], and Steven G. Louie[1,2*]

**Affiliations:**

[1]Department of Physics, University of California at Berkeley, Berkeley, CA, USA.

[2]Materials Sciences Division, Lawrence Berkeley National Laboratory, Berkeley, CA, USA.

*Corresponding author: sglouie@berkeley.edu (S.G.L.)

## Abstract

Combining magnetometry with optical spectroscopy has uncovered novel quantum phenomena and is emerging as a powerful probe of quantum materials. However, the theory of the magnetic response of excitons, correlated electron-hole pairs in insulators, remains incomplete due to insufficient treatment of electron-hole interactions and quantum geometric effects. In biased bilayer graphene, for instance, theoretical predictions of valley g-factors for p-excitons deviate from experiment by nearly an order of magnitude. Here, we develop a quantum theory of the exciton orbital magnetic moment, based on first-order perturbation theory within the GW plus Bethe-Salpeter Equation approach and a rigorous treatment of the position operator in the response of exciton states to a magnetic field. Our formalism reveals three distinct contributions that go beyond the heuristic approaches used in the literature: a Berry-phase-corrected single-particle electron and hole moment difference, a term from envelope-function winding linked to electron-hole relative motion, and a center-of-mass correction from exciton band quantum geometry, with the latter two being completely new effects not considered in previous studies. Our ab initio calculations yield results in excellent agreement with experiment, establishing the importance of interaction and quantum geometric effects in the magnetic response of excitons.

Using light to probe quantum materials has elucidated a wide range of phenomena, from energy gaps (*1*) and excitonic resonances (*2*) to collective excitations (*3*). More recently, combining optics with magnetometry has emerged as a powerful probe of correlated and topological matter, with the magnetic response of optical excitations providing access to magnetic order (*4*), optical control of magnetism (*5, 6*), and underlying quantum geometry (*7, 8*). Central to these phenomena are excitons, correlated electron-hole (e-h) pairs that dominate the optical response of gapped systems, particularly in low-dimensional materials (*9*). The response of excitons to external magnetic fields in two-dimensional (2D) semiconductors has attracted significant attention, in particular, due to its relevance to valleytronics. For example, in 2D systems with three-fold rotational symmetry and broken inversion symmetry, such as transition metal dichalcogenides (*10-13*) and biased multilayer graphene (*14*), valley degrees of freedom couple to external magnetic fields via Berry curvature (*7, 15, 16*), giving rise to measurable Zeeman splitting of excitonic resonances, commonly quantified by an effective exciton valley g-factor.

Despite extensive experimental progress, a rigorous quantum mechanical theory of the response of excitons to an external magnetic field remains incomplete. To first order, the energy shift of a particle is proportional to its magnetic moment, including both spin and orbital contributions. Conventional approaches approximate the exciton orbital magnetic moment either as the difference of single-particle (band-state) electron and hole contributions (*17-19*) at a critical **k** point or as an average of band-state contributions over the Brillouin-zone (BZ) weighted by the **k**-space exciton envelope function (*20*). These treatments neglect electron-hole interactions, internal exciton structure, and center-of-mass (COM) dynamics, and fail in systems such as biased bilayer graphene, where predicted p-exciton g-factors deviate from experiment by nearly an order of magnitude (*14*). The underlying difficulty lies in the proper and full treatment of the orbital magnetic moment operator which involves the position operator matrix elements in extended Bloch states basis (*21-23*). These matrix elements contain singular terms proportional to $\nabla_{\mathbf{k}}\delta(\mathbf{k}-\mathbf{k}')$ and neglecting these leads to incomplete and gauge-dependent results. The issue is further amplified for excitons, which are composite quasiparticles with both internal and COM Bloch-state character.

In this work, we develop a quantum mechanical theory of the exciton orbital magnetic moment within the many-body GW plus Bethe-Salpeter equation (GW-BSE) approach (*24*, *25*). By introducing a periodic analogue of the position operator and constructing localized exciton wave packets in the COM coordinate, we resolve the divergences of the position operator in Bloch basis, and obtain, in the appropriate limits, a gauge-invariant expression for the exciton orbital magnetic moment. A schematic of our approach is shown in Fig. 1a. We show that the exciton orbital magnetic moment naturally decomposes into contributions from Berry-phase-corrected single-particle moments, winding of the exciton envelope function, and a previously overlooked term arising from exciton band quantum geometry (Fig. 1b). We compute the exciton valley g-factors for *s*- and *p*-type excitons in biased bilayer graphene (BBG), and the results show excellent agreement with experiment, resolving the discrepancies exhibited in prior theories.

To determine the orbital magnetic moment of an exciton state with quantum number $n$ and COM momentum $\mathbf{Q}$, $\boldsymbol{\mu}^X_{n\mathbf{Q}}$, we consider its first-order change in energy $\delta E^X_{n\mathbf{Q}}$ in the presence of an uniform magnetic field $\mathbf{B}$, $\delta E^X_{n\mathbf{Q}} = -\boldsymbol{\mu}^X_{n\mathbf{Q}} \cdot \mathbf{B}$ (see Supplementary Material (SM) Sec. 1 (*26*)). We formulate this problem starting with the unperturbed exciton Hamiltonian, $H^X$, within a two-particle framework such as the GW-BSE approach (*25*). Owing to translation symmetry, the COM momentum $\mathbf{Q}$ is a good quantum number, leading to $H^X = \sum_{\mathbf{Q}} H^X(\mathbf{Q})$. In the free e-h band state basis, $|vc\mathbf{kQ}\rangle = \left|c\mathbf{k}+\frac{\mathbf{Q}}{2}\right\rangle \otimes \left|v\mathbf{k}-\frac{\mathbf{Q}}{2}\right\rangle^*$, the Hamiltonian reads:

$$
\begin{aligned}
H^X_{vc\mathbf{k},v'c'\mathbf{k}'}(\mathbf{Q}) &= \langle vc\mathbf{kQ}|H^X|v'c'\mathbf{k}'\mathbf{Q}\rangle \\
&= \left(\epsilon_{c\mathbf{k}+\mathbf{Q}/2} - \epsilon_{v\mathbf{k}-\mathbf{Q}/2}\right)\delta_{vv'}\delta_{cc'}\delta_{\mathbf{kk}'} + K^{eh}_{vc\mathbf{k},v'c'\mathbf{k}'}(\mathbf{Q}),
\end{aligned} \tag{1}
$$

where $K^{eh}$ is the e-h interaction kernel. Diagonalizing $H^X(\mathbf{Q})$ yields exciton energies $E^X_{n\mathbf{Q}}$ (the exciton band structure) and eigenvectors $A^{n\mathbf{Q}}_{vc\mathbf{k}}$ (called the exciton envelope function), with corresponding exciton states $|X_{n\mathbf{Q}}\rangle = \sum_{vc\mathbf{k}} A^{n\mathbf{Q}}_{vc\mathbf{k}}|vc\mathbf{kQ}\rangle$ written in the electron-hole coordinate $(\mathbf{r_e}, \mathbf{r_h})$ representation as,

$$
\begin{aligned}
\langle \mathbf{r_e}, \mathbf{r_h}|X_{n\mathbf{Q}}\rangle &= e^{i\mathbf{Q}\cdot\left(\frac{\mathbf{r_e}+\mathbf{r_h}}{2}\right)} \sum_{vc\mathbf{k}} A^{n\mathbf{Q}}_{vc\mathbf{k}} e^{i\mathbf{k}\cdot(\mathbf{r_e}-\mathbf{r_h})} u_{c\mathbf{k}+\mathbf{Q}/2}(\mathbf{r_e}) u^*_{v\mathbf{k}-\mathbf{Q}/2}(\mathbf{r_h}) \\
&= e^{i\mathbf{Q}\cdot\left(\frac{\mathbf{r_e}+\mathbf{r_h}}{2}\right)} \langle \mathbf{r_e}, \mathbf{r_h}|u^X_{n\mathbf{Q}}\rangle = e^{i\mathbf{Q}\cdot\left(\frac{\mathbf{r_e}+\mathbf{r_h}}{2}\right)} u^X_{n\mathbf{Q}}(\mathbf{r_e}, \mathbf{r_h}),
\end{aligned} \tag{2}
$$

where $|u^X_{n\mathbf{Q}}\rangle$ is the periodic part of the exciton Bloch wavefunction in the COM coordinate $\mathbf{R} = \frac{\mathbf{r_e}+\mathbf{r_h}}{2}$. The magnetic response is obtained from matrix elements of the perturbed Hamiltonian $\widetilde{H}^X$ in the *unperturbed* excitonic state basis $|X_{n\mathbf{Q}}\rangle$, i.e., by evaluating $\langle X_{n\mathbf{Q}}|\widetilde{H}^X|X_{n'\mathbf{Q}'}\rangle$. The quasiparticle part of the GW-BSE Hamiltonian is treated perturbatively to first order in the magnetic field. For evaluating the field-dependent e-h interaction kernel $K^{eh}_{vc\mathbf{k},v'c'\mathbf{k}'}(\mathbf{Q},\mathbf{B})$, we similarly use quasiparticle wavefunctions to first-order in $\mathbf{B}$; but we adopt a fixed-screening approximation, whereby the screened Coulomb interaction $W$ that enters $K^{eh}_{vc\mathbf{k},v'c'\mathbf{k}'}$ is kept at zero magnetic field, i.e., $\widetilde{W} \approx W$ (see SM Sec. 2 (*26*)). Consequently, the magnetic-field dependence of the screened Coulomb interaction, which only induces numerically small corrections, is neglected, while the first-order magnetic response in $K^{eh}_{vc\mathbf{k},v'c'\mathbf{k}'}(\mathbf{Q},\mathbf{B})$ arising from the changes in the quasiparticle energies and wavefunctions are treated explicitly. We hence obtain:

$$\begin{aligned}\langle X_{n\mathbf{Q}}|\widetilde{H}^X|X_{n'\mathbf{Q}'}\rangle = \delta_{\mathbf{Q},\mathbf{Q}'}\delta_{n,n'}\epsilon^X_{n\mathbf{Q}} - \\ \mathbf{B}\cdot\sum_{vc\mathbf{k}}\Big[\sum_{c'\mathbf{k}'} A^{n\mathbf{Q}*}_{vc\mathbf{k}}A^{n'\mathbf{Q}'}_{vc'\mathbf{k}'}\delta_{\mathbf{k}-\frac{\mathbf{Q}}{2},\mathbf{k}'-\frac{\mathbf{Q}'}{2}}\left\langle c\mathbf{k}+\frac{\mathbf{Q}}{2}\middle|\hat{\boldsymbol{\mu}}\middle|c'\mathbf{k}'+\frac{\mathbf{Q}'}{2}\right\rangle \\ -\sum_{v'\mathbf{k}'} A^{n\mathbf{Q}*}_{vc\mathbf{k}}A^{n'\mathbf{Q}'}_{v'c\mathbf{k}'}\delta_{\mathbf{k}+\frac{\mathbf{Q}}{2},\mathbf{k}'+\frac{\mathbf{Q}'}{2}}\left\langle v'\mathbf{k}'-\frac{\mathbf{Q}'}{2}\middle|\hat{\boldsymbol{\mu}}\middle|v\mathbf{k}-\frac{\mathbf{Q}}{2}\right\rangle\Big].\end{aligned} \tag{3}$$

The key challenge lies in evaluating matrix elements of the orbital magnetic moment operator $\hat{\boldsymbol{\mu}} = -\frac{e}{2}\hat{\mathbf{r}}\times\hat{\mathbf{v}}$ in the single-particle Bloch basis, where the intra-band electron position operator matrix element introduces singular terms proportional to $\nabla_{\mathbf{k}}\delta(\mathbf{k}-\mathbf{k}')$ (see SM Sec. 3 (*26*)). Previous treatments effectively neglect these contributions, leading to incomplete formalism (*17-20, 27*). To address this, we introduce a periodic representation of the position operator $\hat{\mathbf{r}} \rightarrow \hat{\mathbf{r}}_{\boldsymbol{\lambda}} = -i\nabla_{\boldsymbol{\lambda}}e^{i\boldsymbol{\lambda}\cdot\hat{\mathbf{r}}}$, which is well-defined in the Bloch basis and recovers the physical operator $\hat{\mathbf{r}}$ in the long-wavelength limit ($\boldsymbol{\lambda}\rightarrow 0$). This corresponds to evaluating the response to a spatially periodic magnetic field and then taking the uniform limit (*28, 29*). Following this procedure, we obtain the perturbed exciton Hamiltonian to first-order in the $\mathbf{B}$ field, as presented in SM Sec. 3 (*26*). There appears a term proportional to $\nabla_{\mathbf{Q}}\delta_{\mathbf{Q},\mathbf{Q}'}$, which reflects the extended nature of exciton Bloch states and requires a careful treatment (*30, 31*). To address this rigorously, we construct a localized exciton wave packet $|W_{n\mathbf{Q}_c}\rangle = \sum_{\mathbf{Q}} w(\mathbf{Q}-\mathbf{Q}_c)|X_{n\mathbf{Q}}\rangle$, centered at $\mathbf{Q}_c$, and evaluate its energy shift before taking the infinite-extent limit $|w(\mathbf{Q}-\mathbf{Q}_c)|^2 \rightarrow \delta(\mathbf{Q}-\mathbf{Q}_c)$ (see SM Sect. 4 (*26*)). This procedure yields the physical result and leads to a gauge-invariant expression (see SM Sect. 5 (*26*)) for the exciton magnetic moment:

$$\begin{aligned}\boldsymbol{\mu}^X_{n\mathbf{Q}} = \sum_{vc\mathbf{k}}\left\{\sum_{c'} A^{n\mathbf{Q}*}_{vc\mathbf{k}}A^{n\mathbf{Q}}_{vc'\mathbf{k}}\boldsymbol{\mu}_{cc'\mathbf{k}+\frac{\mathbf{Q}}{2}} - \sum_{v'} A^{n\mathbf{Q}*}_{vc\mathbf{k}}A^{n\mathbf{Q}}_{v'c\mathbf{k}}\boldsymbol{\mu}'_{v'v\mathbf{k}-\frac{\mathbf{Q}}{2}}\right\} \\ +\frac{e}{2\hbar}\sum_{vc\mathbf{k}}\left\{\sum_{c'} A^{n\mathbf{Q}*}_{vc\mathbf{k}}A^{n\mathbf{Q}}_{vc'\mathbf{k}}\left(\nabla_{\mathbf{k}}\epsilon_{c\mathbf{k}}|_{\mathbf{k}+\frac{\mathbf{Q}}{2}}\times\boldsymbol{\mathcal{A}}_{cc'\mathbf{k}+\frac{\mathbf{Q}}{2}}\right) - \sum_{v'} A^{n\mathbf{Q}*}_{vc\mathbf{k}}A^{n\mathbf{Q}}_{v'c\mathbf{k}}\left(\nabla_{\mathbf{k}}\epsilon_{v\mathbf{k}}|_{\mathbf{k}-\frac{\mathbf{Q}}{2}}\times\boldsymbol{\mathcal{A}}_{v'v\mathbf{k}-\frac{\mathbf{Q}}{2}}\right)\right\} \\ +\frac{e}{4i}\sum_{vc\mathbf{k}}\left\{\sum_{c'}\left(A^{n\mathbf{Q}*}_{vc\mathbf{k}}\nabla_{\mathbf{k}}A^{n\mathbf{Q}}_{vc'\mathbf{k}}\times\mathbf{v}_{cc'\mathbf{k}+\frac{\mathbf{Q}}{2}}\right) + \sum_{v'}\left(A^{n\mathbf{Q}*}_{vc\mathbf{k}}\nabla_{\mathbf{k}}A^{n\mathbf{Q}}_{v'c\mathbf{k}}\times\mathbf{v}_{v'v\mathbf{k}-\frac{\mathbf{Q}}{2}}\right)\right\} \\ +\frac{e}{2i}\sum_{vc\mathbf{k}}\left\{\sum_{c'}\left(A^{n\mathbf{Q}*}_{vc\mathbf{k}}\mathbf{D}_{\mathbf{Q}}A^{n\mathbf{Q}}_{vc'\mathbf{k}}\times\mathbf{v}_{cc'\mathbf{k}+\frac{\mathbf{Q}}{2}}\right) - \sum_{v'}\left(A^{n\mathbf{Q}*}_{vc\mathbf{k}}\mathbf{D}_{\mathbf{Q}}A^{n\mathbf{Q}}_{v'c\mathbf{k}}\times\mathbf{v}_{v'v\mathbf{k}-\frac{\mathbf{Q}}{2}}\right)\right\},\end{aligned} \tag{4}$$

where the notations used are defined as follows: We express the matrix elements of the single-particle orbital magnetic moment operator in the multi-band case using two matrices: $\boldsymbol{\mu}_{ij\mathbf{k}} = \frac{e}{2i\hbar}\langle\nabla_{\mathbf{k}}u_{i\mathbf{k}}|\times[H_{\mathbf{k}}-\epsilon_{i\mathbf{k}}]|\nabla_{\mathbf{k}}u_{j\mathbf{k}}\rangle$, and $\boldsymbol{\mu}'_{ij\mathbf{k}} = \frac{e}{2i\hbar}\langle\nabla_{\mathbf{k}}u_{i\mathbf{k}}|\times[H_{\mathbf{k}}-\epsilon_{j\mathbf{k}}]|\nabla_{\mathbf{k}}u_{j\mathbf{k}}\rangle$ with $H_{\mathbf{k}} = e^{-i\mathbf{k}\cdot\mathbf{r}}He^{i\mathbf{k}\cdot\mathbf{r}}$. Note that the two matrices $\boldsymbol{\mu}_{ij\mathbf{k}}$, and $\boldsymbol{\mu}'_{ij\mathbf{k}}$ are different with the placement of $\epsilon_{i\mathbf{k}}$ or $\epsilon_{j\mathbf{k}}$ that goes inside the expression, respectively. The single-particle inter-band Berry connection is given by: $\boldsymbol{\mathcal{A}}_{ij\mathbf{k}} = i\langle u_{i\mathbf{k}}|\nabla_{\mathbf{k}}u_{j\mathbf{k}}\rangle$, and the velocity matrix by: $\mathbf{v}_{ij\mathbf{k}} = \langle u_{i\mathbf{k}}|\hat{\mathbf{v}}(\mathbf{k})|u_{j\mathbf{k}}\rangle = \frac{1}{\hbar}(\epsilon_{j\mathbf{k}}-\epsilon_{i\mathbf{k}})\langle u_{i\mathbf{k}}|\nabla_{\mathbf{k}}u_{j\mathbf{k}}\rangle + \frac{1}{\hbar}\delta_{ij}\nabla_{\mathbf{k}}\epsilon_{i\mathbf{k}}$. We also define a covariant derivative with respect to exciton COM momentum given by $\mathbf{D}_{\mathbf{Q}} = \nabla_{\mathbf{Q}} + i\boldsymbol{\mathcal{A}}^{\mathrm{X}}_{n\mathbf{Q}}$ with $\boldsymbol{\mathcal{A}}^{\mathrm{X}}_{n\mathbf{Q}} = i\langle u^X_{n\mathbf{Q}}|\nabla_{\mathbf{Q}}u^X_{n\mathbf{Q}}\rangle$ being the exciton band Berry connection.

Eq. 4 contains four terms, which taken together is gauge invariant. The first term represents a multiband generalization of previously proposed expressions for the exciton orbital magnetic moment (*20, 27*). By itself, however, this term is not gauge-invariant. Previous multiband formulations (arriving at approximate expressions related to the physics of this term) obtain gauge-invariant expressions by introducing additional approximations in the treatment of the position operator in the Bloch basis, namely by neglecting either the intra-band or the inter-band matrix elements (see SM Sec. 6 (*26*)). In contrast, Eq. 4 is obtained by rigorously treating both the intra-band and inter-band matrix elements of the position operator, resulting in a fully gauge-invariant expression that naturally captures the additional relative-motion and COM quantum-geometric contributions absent in previous approximations. For physical interpretation, we regroup the terms into three physically distinct contributions:

$$\boldsymbol{\mu}_{n\mathbf{Q}}^{X} = \boldsymbol{\mu}_{n\mathbf{Q}}^{X,sp} + \boldsymbol{\mu}_{n\mathbf{Q}}^{X,rel} + \boldsymbol{\mu}_{n\mathbf{Q}}^{X,COM}, \tag{5}$$

where $\boldsymbol{\mu}_{n\mathbf{Q}}^{X,sp}$ corresponds to the Berry-phase-corrected single-particle ($sp$) contributions (sum of terms #1 and #2 in Eq. 4), $\boldsymbol{\mu}_{n\mathbf{Q}}^{X,rel}$ arises from the relative ($rel$) e-h motion (term #3 in Eq. 4), and $\boldsymbol{\mu}_{n\mathbf{Q}}^{X,COM}$ originates from the exciton $COM$ band quantum geometry (term #4 in Eq. 4). Further insights to the physical interpretation of these contributions within the two-band limit are discussed in the end matter.

We next compute the exciton orbital magnetic moment (the full Eqs. 4 and 5) in biased bilayer graphene (BBG), where exciton valley g-factors have been experimentally measured (*14*). We perform DFT calculations including hBN encapsulation (Fig. 2a), followed by GW (*24*) and GW-BSE (*25*) calculations for quasiparticle bands and excitons (see SM Sec. 7 (*26*)). For an applied field of 0.10 eV/Å, the quasiparticle gap is 161 meV, significantly enhanced over DFT. The bands near K and K′ exhibit a Mexican-hat dispersion (Fig. 2b), with a clear particle-hole asymmetry, leading to a finite difference in single-particle orbital magnetic moments (Fig. 2c) (also see SM Sec. 8 (*26*)).

As seen in previous theoretical (*32*) and experimental studies (*14*), the optical absorption spectrum (Fig. 3a) shows two bright excitons below the gap (*32*), corresponding to *s*- and *p*-type excitons at $\mathbf{Q} = 0$. Their envelope functions (Fig. 3b, 3c) display distinct phase structures: the *s*-exciton is node-less, while the *p* exciton exhibits a $2\pi$ phase winding and a node, consistent with angular momentum $+1$. The exciton dispersion (Fig. 4a) shows exchange-induced splitting at finite $\mathbf{Q}$, with a parabolic and a nonanalytic branch (*33-36*). This splitting is larger for *p*-exciton due to its larger oscillator strength. Furthermore, the excitonic bands carry a pseudospin texture inherited from their valley composition (*33*) exhibiting winding $\pm 2$ (Fig. 4b).

These properties make BBG an ideal platform to probe the contributions in Eq. 5, including those from envelope-function phase winding and exciton-band quantum geometry. Recent photocurrent spectroscopy measurements on BBG (*14*) has reported the valley g-factor of these two excitons, defined as the difference in orbital magnetic moments between excitons in opposite valleys normalized by the Bohr Magneton ($\mu_B$) with values $g_s = 19.8 \pm 0.1$, and $g_p = 1.4 \pm 0.8$ (Table 1), providing a stringent test of our theory. These experimental values are extracted from photocurrent peak splitting (*14*); the $g_s$ uncertainty reflects fitting error, while the $g_p$ value is inferred based on experimentalists' estimate that the splitting is 20% of the linewidth with uncertainties derived from the linewidth variation across magnetic fields.

At the simplest level, a common approximation (*17, 18*) in the literature is to estimate the exciton orbital magnetic moment from the band-edge difference of single-particle moments, neglecting the exciton wavefunction and Berry-curvature effects. This yields a valley g-factor for the bright (like-spin transitions) excitons with zero spin contribution:

$$g = \frac{1}{\mu_B} |[\boldsymbol{\mu}_{c\mathbf{k}=\mathrm{K}} - \boldsymbol{\mu}_{v\mathbf{k}=\mathrm{K}}]_z - [\boldsymbol{\mu}_{c\mathbf{k}=\mathrm{K}'} - \boldsymbol{\mu}_{v\mathbf{k}=\mathrm{K}'}]_z|, \quad (6)$$

giving g = 15.06 for both *s*- and *p*-excitons for our BBG, in clear disagreement with experiment at the bias electric field considered. $\boldsymbol{\mu}_{n\mathbf{k}}$ is calculated using sum-over-states method, as discussed in SM Sec. 8 (*26*). A more refined, but heuristic, approach (*20*) includes the exciton envelope function via a momentum-space average of the single-particle moment difference (first term on RHS of Eq. 4), leading to

$$g_n = \frac{1}{\mu_B} \left| \sum_{\mathbf{k}\in\mathrm{K}} \left|A_{\mathbf{k}}^{n,\mathbf{Q}=\mathbf{0}}\right|^2 [\boldsymbol{\mu}_{c\mathbf{k}} - \boldsymbol{\mu}_{v\mathbf{k}}]_z - \sum_{\mathbf{k}'\in\mathrm{K}'} \left|A_{\mathbf{k}'}^{n,\mathbf{Q}=\mathbf{0}}\right|^2 [\boldsymbol{\mu}_{c\mathbf{k}'} - \boldsymbol{\mu}_{v\mathbf{k}'}]_z \right|. \quad (7)$$

Here the summation over $\mathbf{k}$ ($\mathbf{k}'$) goes over the points near the K (K′) valley. The *s*-exciton, whose envelope function is node-less and broadly distributed in $\mathbf{k}$-space, yields a larger g-factor of 18.45 in this approximation. In contrast, the *p*-exciton's nodal structure suppresses contributions from the regions of maximal $[\boldsymbol{\mu}_{c\mathbf{k}} - \boldsymbol{\mu}_{v\mathbf{k}}]_z$ (Fig. 2c), reducing its g-factor to 13.03. This approach still fails to reproduce the strong suppression of the *p*-exciton g-factor (Table 1).

Using our derived expression (Eqs. 4 and 5), we compute the valley g-factor from the splitting of the two valley-related exciton states $n_1$ and $n_2$ at $\mathbf{Q} = 0$,

$$g_n = \frac{1}{\mu_B} \left| \left[\boldsymbol{\mu}^X_{n_2,\mathbf{Q}=0} - \boldsymbol{\mu}^X_{n_1,\mathbf{Q}=0}\right]_z \right|, \quad (8)$$

where $\boldsymbol{\mu}^X_{n_1,\mathbf{Q}=0}$ and $\boldsymbol{\mu}^X_{n_2,\mathbf{Q}=0}$ are the exciton orbital magnetic moment of the two valley-related exciton band states (as shown in Fig. 4a), evaluated at COM momentum $\mathbf{Q} = 0$. Using this approach, we obtain $g_s = 20.32$ and $g_p = 1.79$, in excellent agreement with experiment (Table 1) and significantly improved over those from previous approximations due to interaction and quantum geometric effects overlooked in prior theories. Since the exciton states in BBG are overwhelmingly dominated by transitions from the highest valence band to the lowest conduction band (see SM Sec. 7 (*26*)), we have also evaluated the exciton g-factors using the exciton orbital moments calculated using the approximate two-band expression (sum of Eqs. 9, 10, and 11 in End Matter below). The resulting values, $g_s = 20.25$ and $g_p = 1.93$, differ only slightly from the full multi-band results (Table 1), demonstrating that the two-band approximation captures the dominant physics in this particular material while the full multi-band formalism provides a small but finite correction. We expect multi-band effects to become considerably more important in systems where the exciton wavefunction has significant contributions from multiple valence-conduction pairs, such as moiré materials (*37*) or in general systems with degenerate or nearly degenerate valence or conduction manifolds. The present multi-band formalism naturally captures such effects.

To elucidate the origin of the contrasting g-factors for the *s*- and *p*-excitons in BBG, we decompose the exciton magnetic moment ($\boldsymbol{\mu}^X_{s_1,\mathbf{Q}=0}$ and $\boldsymbol{\mu}^X_{p_1,\mathbf{Q}=0}$) into its constituent contributions (Table 2). Note that the individual terms in Table 2 are gauge dependent, but the sum which is the physical moment is gauge invariant. The following discussion is based on a gauge choice in which the lowest s-exciton envelope function has a uniform phase (*38*). Since the excitons are degenerate at $\mathbf{Q} = 0$, to evaluate the $\mathbf{Q}$-covariant derivatives we further choose a basis within the degenerate subspace in which the two exciton envelope functions are localized in opposite valleys. The Berry-phase correction to the single-particle term is substantial, contributing +2.23 (*s*) and +3.32 (*p*), and is enhanced for the *p*-exciton due to its envelope function localization in regions of large electron and hole Berry curvature (see SM Fig. S4 (*26*)). This contribution reflects the orbital magnetic moments of the underlying conduction and valence band states and is therefore similar for both *s*- and *p*-type excitons, apart from the envelope-function averaging.

The relative-motion contribution is negligible for the *s*-exciton but large and negative for the *p*-exciton ($-5.95$), reflecting the phase winding of the *p*-exciton envelope function. The *s*-exciton has no envelope-function winding since there is no angular momentum associated with the e-h relative motion. In contrast, for it to be optically active, the phase winding the envelope function of the *p*-exciton in $\mathbf{k}$-space is constrained by the winding-number optical selection rule (32) to compensate the winding of the inter-band transition dipole matrix elements. Since the latter is determined by the chirality of the underlying conduction and valence Bloch states, the relative-motion contribution necessarily for the optically active p-exciton acquires the opposite sign to the single-particle contribution.

The COM contribution shows a similar trend, with a small value for the *s*-exciton ($-1.22$) and a larger magnitude for the *p*-exciton ($-4.78$). This contribution originates from the quantum geometry of the exciton band structure and the exchange-driven valley pseudospin texture. Within the effective $\mathbf{Q} \cdot \mathbf{p}$ description (33), the inter-valley exchange matrix element is proportional to the optical transition dipole moment $\langle 0|\hat{\mathbf{p}}|X_{\mathbf{n0}}\rangle$, whose phase winding follows that of the exciton envelope function. Consequently, the COM pseudospin winding has the same chirality as the envelope-function winding and therefore contributes with the same sign as the relative-motion term discussed above. Because the *p*-exciton is optically brighter and undergoes much stronger inter-valley exchange mixing, this quantum-geometric contribution is substantially larger than that for the *s*-exciton.

As a result, the two contributions, the relative-motion angular momentum and the COM quantum geometry, both oppose the single-particle contribution, leading to a near cancellation for the *p*-exciton and hence its very small magnetic moment. In contrast, the *s*-exciton possesses neither relative-motion angular momentum nor significant exchange-induced COM pseudospin, so its magnetic moment remains dominated by the single-particle contribution. The strong suppression of the *p*-exciton g-factor is therefore a direct consequence of the symmetry and chirality of the exciton wavefunction rather than an accidental cancellation. This highlights the essential roles of exciton quantum geometry and envelope-function structure in determining the magnetic response of excitons, effects that are absent in previous theoretical treatments.

In summary, we have derived, within the GW-BSE formalism (25), a complete quantum-mechanical expression for the exciton orbital magnetic moment by rigorously treating the position operator in the Bloch basis and incorporating electron-hole interaction and quantum-geometric effects. The resulting gauge-invariant expression naturally separates the exciton magnetic moment into Berry-phase-corrected single-particle, relative-motion, and COM quantum-geometric contributions. Applying this theory to biased bilayer graphene, we obtain exciton valley g-factors in quantitative agreement with experiment, resolving the discrepancy with previous theoretical approaches and revealing the physical origin of the strongly suppressed *p*-exciton magnetic moment. Our formalism is general for any periodic crystalline materials and is not restricted to valley systems or two-dimensional materials. For three-dimensional crystals, Eq. 4 is directly applicable for magnetic fields applied along any direction. For two-dimensional materials, the present implementation applies to out-of-plane magnetic fields. The extension to in-plane magnetic fields is straightforward, which requires a careful treatment of the out-of-plane position operator and will be presented elsewhere.

The methodology developed here is also not limited to the particle-hole BSE formalism. The same perturbative framework, based on constructing the magnetic-field-perturbed two-particle Hamiltonian in the unperturbed correlated two-particle state basis together with the periodic treatment of the position operator, can be extended to other two-particle channels, e.g. using a particle-particle BSE formalism (39). Finally, the present framework can be naturally extended to incorporate full-spinor wavefunctions, the full electron-hole interaction kernel beyond the Tamm-Dancoff approximation (25), and systems with degenerate bands, where non-Abelian quantum geometric contributions may arise. Extensions to higher-order perturbations will enable studies of diamagnetic and nonlinear responses, providing a general platform for exciton magnetism and quantum geometry across diverse materials, including multi-layer systems (*40*), moiré heterostructures (*37*), and topological flat band materials (*41*).

References in Supplementary Material: (42-51)

## End Matter

**Physical interpretation of the different contributions to the exciton orbital magnetic moment in the two-band limit:**

To gain some physical insight into the various contributions to the exciton orbital magnetic moment (Eq. 4 and 5), we consider the two-band limit involving a single valence and conduction band. In this case, the three contributions in Eq. 5 admit simpler expressions and provide physical interpretations.

The single-particle contribution reduces to

$$\boldsymbol{\mu}_{n\mathbf{Q}}^{X,sp}(2band) = \sum_{\mathbf{k}} \left|A_{\mathbf{k}}^{n\mathbf{Q}}\right|^2 \left(\boldsymbol{\mu}_{c\mathbf{k}+\frac{\mathbf{Q}}{2}} - \boldsymbol{\mu}_{v\mathbf{k}-\frac{\mathbf{Q}}{2}}\right) + \frac{e}{2\hbar}\sum_{\mathbf{k}} \left|A_{\mathbf{k}}^{n\mathbf{Q}}\right|^2 \left(\nabla_{\mathbf{k}}\epsilon_{c\mathbf{k}}\big|_{\mathbf{k}+\frac{\mathbf{Q}}{2}} \times \boldsymbol{\mathcal{A}}_{c\mathbf{k}+\frac{\mathbf{Q}}{2}} - \nabla_{\mathbf{k}}\epsilon_{v\mathbf{k}}\big|_{\mathbf{k}-\frac{\mathbf{Q}}{2}} \times \boldsymbol{\mathcal{A}}_{v\mathbf{k}-\frac{\mathbf{Q}}{2}}\right). \tag{9}$$

The first term on the RHS of Eq. 9 corresponds to the difference between the quasiparticle electron and hole orbital magnetic moments averaged over **k**-space with the exciton envelope function amplitude squared, which is the heuristic expression typically used in the literature so far (20). The second term on the RHS represents a Berry-phase correction, arising from our rigorous treatment of the position operator, due to the modified density of states in the presence of external field (52). This correction is absent in previous approaches even at the single-particle level.

The relative-motion contribution reduces to

$$\boldsymbol{\mu}_{s\mathbf{Q}}^{X,rel}(2band) = \frac{e}{4i\hbar}\sum_{\mathbf{k}} A_{\mathbf{k}}^{n\mathbf{Q}*}\nabla_{\mathbf{k}}A_{\mathbf{k}}^{n\mathbf{Q}} \times \left(\nabla_{\mathbf{k}}\epsilon_{c\mathbf{k}}\big|_{\mathbf{k}+\frac{\mathbf{Q}}{2}} + \nabla_{\mathbf{k}}\epsilon_{v\mathbf{k}}\big|_{\mathbf{k}-\frac{\mathbf{Q}}{2}}\right), \tag{10}$$

which arises from the internal e-h dynamics. The quantity $A_{\mathbf{k}}^{n\mathbf{Q}*}\nabla_{\mathbf{k}}A_{\mathbf{k}}^{n\mathbf{Q}}$ captures the phase winding of the exciton envelope function in **k**-space, which when multiplied by the relative e-h velocity and reflects the orbital character (e.g., *s*- or *p*-like) of the exciton.

The COM contribution reduces to,

$$\boldsymbol{\mu}_{n\mathbf{Q}}^{X,COM}(2band) = \frac{e}{2i\hbar}\sum_{\mathbf{k}}\left(A_{\mathbf{k}}^{n\mathbf{Q}*}\mathbf{D}_{\mathbf{Q}}A_{\mathbf{k}}^{n\mathbf{Q}}\right) \times \left(\nabla_{\mathbf{k}}\epsilon_{c\mathbf{k}}\big|_{\mathbf{k}+\frac{\mathbf{Q}}{2}} - \nabla_{\mathbf{k}}\epsilon_{v\mathbf{k}}\big|_{\mathbf{k}-\frac{\mathbf{Q}}{2}}\right) \tag{11}$$

It originates from the exciton COM dynamics and reveals a purely quantum effect. This term arises from phase winding of the exciton wavefunction in **Q**-space, while the covariant derivative also involves the exciton-band Berry connection and is linked to the anomalous COM velocity contribution in external fields (*31*). These terms demonstrate that, despite its charge neutrality, the exciton COM motion contributes nontrivially to its orbital magnetic moment through the exciton-band quantum geometry.

**Sensitivity of computed exciton g-factor on $G_0W_0$ starting point:**

We note that the exact values of the computed exciton g-factors depend on the underlying quasiparticle electronic structure, since the exciton magnetic moment (Eq. 4) is given in terms of the quasiparticle energy derivatives, inter-band Berry connections, and velocity matrix elements. Different mean-field starting points for $G_0W_0$ may therefore lead to some variation in the calculated exciton orbital magnetic moments. However, the low-energy electronic structure of BBG is confined to the carbon-$p_z$ manifold (Fig. S1a (*26*)), whose orbital character is expected to be relatively insensitive to the choice of exchange-correlation functional. Consequently, the dependence on the $G_0W_0$ starting point is expected to be weak (See SM Sec. 10 (*26*)). The physical trends and the origin of the different contributions are, moreover, robust against appreciable changes in the underlying electronic structure. This is seen in the calculations for different BBG/hBN structures (see SM Sec. 7 (*26*)), which consistently exhibit a dominant single-particle contribution for the *s*-exciton and strong cancellation between the three contributions for the *p*-exciton. Thus, the decomposition of the exciton magnetic moment into single-particle, relative-motion, and COM quantum-geometric contributions, together with the physical mechanism responsible for the suppression of the *p*-exciton magnetic moment, remains unchanged.

**General convention for exciton wavefunction with finite center-of-mass momentum:**

We further show in the SM Sec. 9 (*26*) that the theory can be formulated using a general convention for the exciton wavefunction with an arbitrary momentum-partition parameter ($\alpha$), i.e., defined using the free e-h pair basis given by $|c, \mathbf{k} + (1-\alpha)\mathbf{Q}\rangle \otimes |v, \mathbf{k} - \alpha\mathbf{Q}\rangle^*$, instead of the symmetric convention used above ($\alpha = 1/2$). (We note that the convention $\alpha = 1/2$ is used for the decomposition presented in Table 2.) This allows the theory to be implemented directly using a different momentum convention adopted by different GW-BSE codes without requiring an explicit transformation to the symmetric convention. We should add however that, with $\mathbf{Q}$ conjugate to $\frac{\mathbf{r_e}+\mathbf{r_h}}{2}$ and $\mathbf{k}$ conjugate to $\mathbf{r_e} - \mathbf{r_h}$ in our symmetric convention, it facilitates the physical interpretation of the derived 3 contributions.

**Acknowledgements:** We thank Xiaoxun Gong, Jack McArthur, and Shang Ren for the fruitful discussions. This work was primarily supported by the Center for Computational Study of Excited-State Phenomena in Energy Materials (C2SEPEM) at LBNL, funded by the U.S. Department of Energy, Office of Science, Basic Energy Sciences, Materials Sciences and Engineering Division under Contract No. DE-AC02-05CH11231, as part of the Computational Materials Sciences Program which provided the developments of the theory, advanced codes and simulations, as well as the *GW* and *GW*-BSE calculations of e-h interactions and exciton properties. Additional support was provided by the National Science Foundation under Grant No. DMR-2325410, which provided topological and gauge invariance analyses and ab initio calculations of exciton magnetic moments. Computational resources were provided by National Energy Research Scientific Computing Center (NERSC), supported by the Office of Science of the U.S. Department of Energy under Contract No. DE-AC02-05CH11231; Stampede2 at the Texas Advanced Computing Center (TACC), the University of Texas at Austin, through Extreme Science and Engineering Discovery Environment (XSEDE), supported by the National Science Foundation under Grant No. ACI-1053575; and Frontera at TACC, supported by the National Science Foundation under Grant No. OAC-1818253.

## Figures

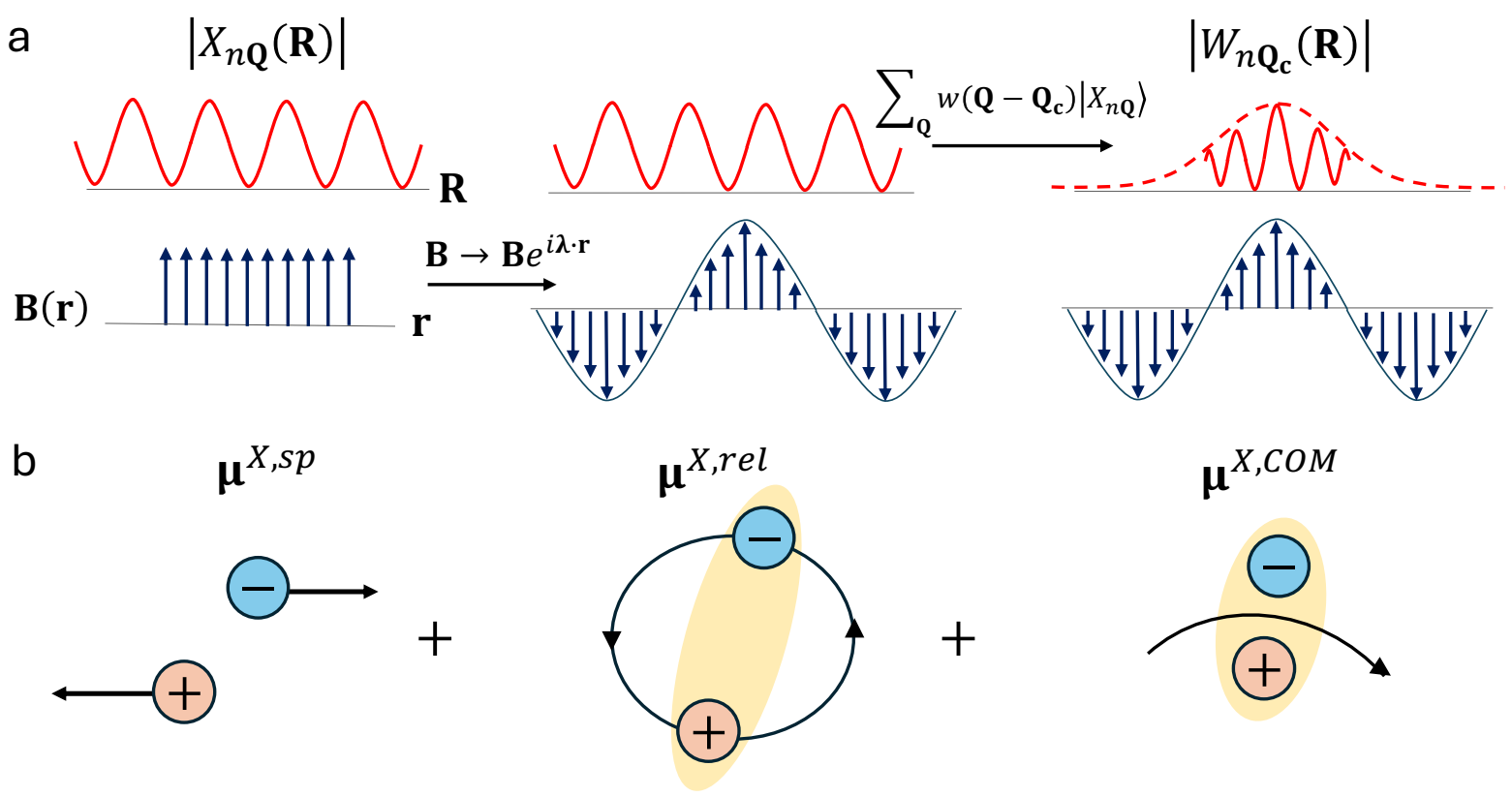

**Fig. 1.** a) Schematic of the derivation for an exciton Bloch state $|X_{n\mathbf{Q}}\rangle$, in a magnetic field. Two divergences arise from the position operator in the Bloch basis: one at the single-particle level (in $\mathbf{k}$ or $\mathbf{r}$) and one from the exciton COM motion (in $\mathbf{Q}$ or $\mathbf{R}$). The $\mathbf{k}$-space divergence is regularized using a periodic representation of the position operator (equivalently, a long-wavelength magnetic field), while the COM divergence is treated by constructing a localized exciton wave packet $|W_{n\mathbf{Q_c}}\rangle$ and taking the delocalization limit. b) The resulting exciton orbital magnetic moment decomposes into three contributions: single-particle ($\boldsymbol{\mu}_{n\mathbf{Q}}^{X,sp}$), relative-motion ($\boldsymbol{\mu}_{n\mathbf{Q}}^{X,rel}$), and COM ($\boldsymbol{\mu}_{n\mathbf{Q}}^{X,COM}$) terms.

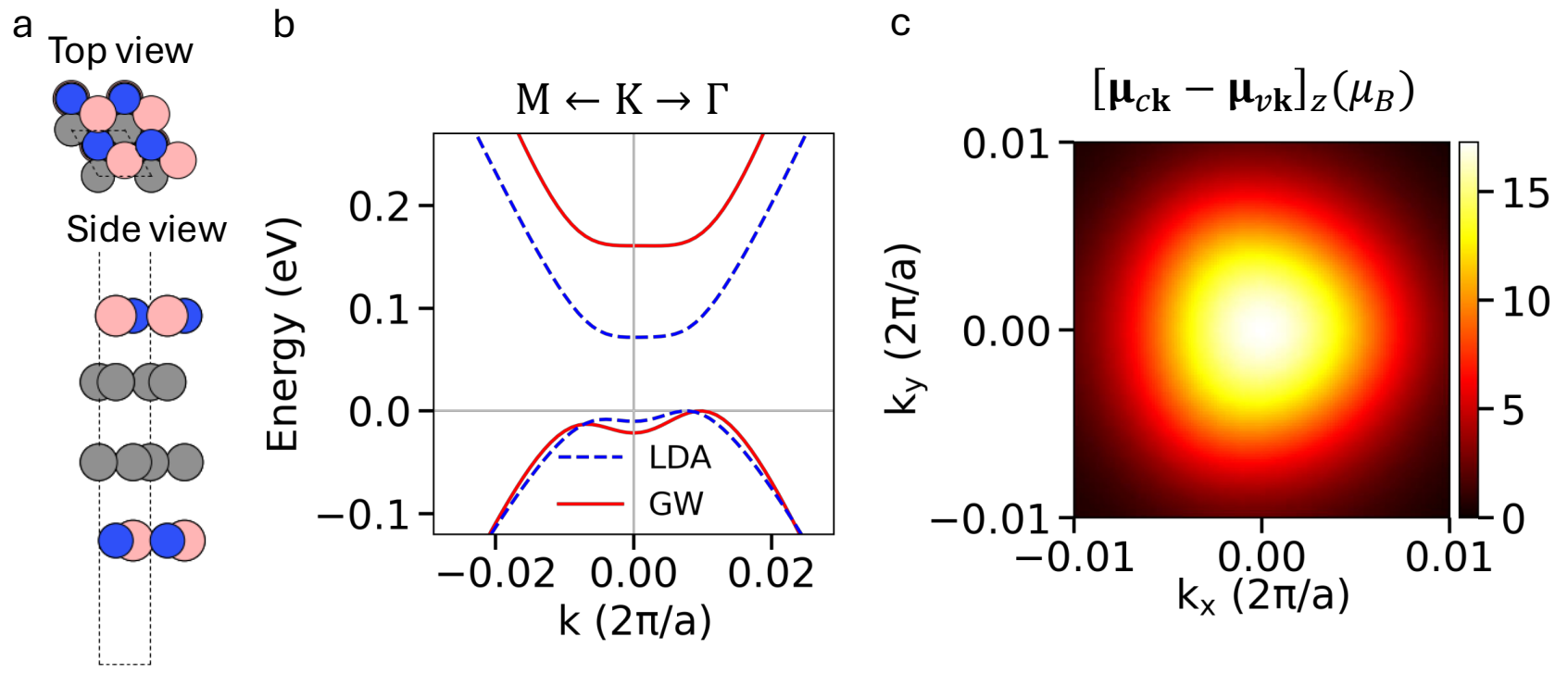

**Fig. 2.** a) Atomic structure of hBN-encapsulated BBG. Grey color balls denote carbon atoms, while blue and pink balls denote nitrogen and boron atoms respectively. b) Electronic band structure near the K valley from DFT-LDA (blue dashed) and GW (red solid) calculations (with bias field of 0.10 eV/Å), showing a significant quasiparticle gap enhancement and modified dispersion. c) Difference between conduction and valence band single-particle orbital magnetic moments near K (in units of $\mu_B$), exhibiting a peak at the valley center and opposite sign at K′ due to time-reversal symmetry.

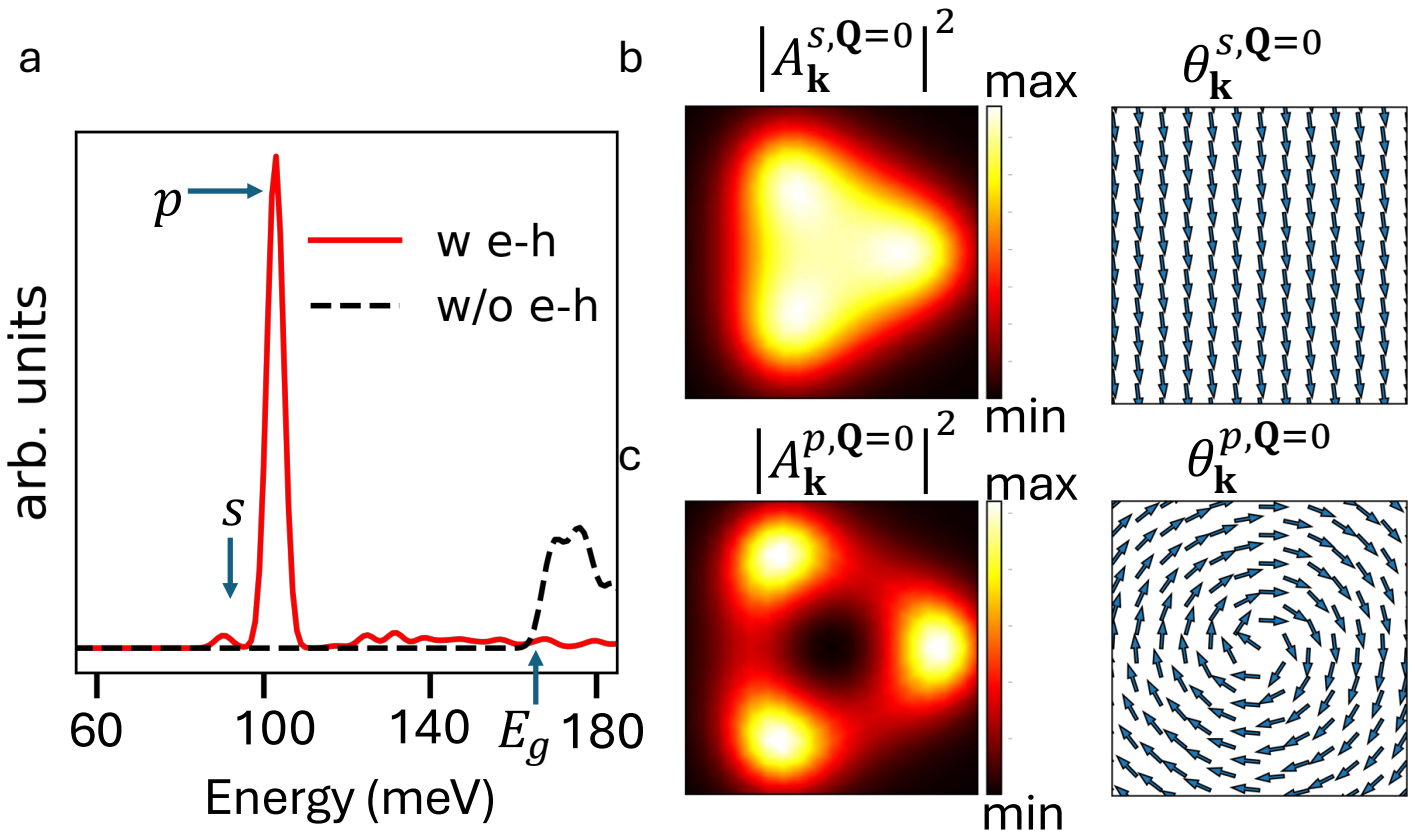

**Fig. 3.** a) Optical absorption with (red) and without (black) e-h interactions, showing bound excitonic peaks below the quasiparticle gap $E_g$. The two lowest peaks correspond to $s$- and $p$-type excitons, each is two-fold degenerate from the valley degree of freedom. b) and c) Momentum-resolved amplitude and phase of the exciton envelope function near K for the $s$ (b) and $p$ (c) states.

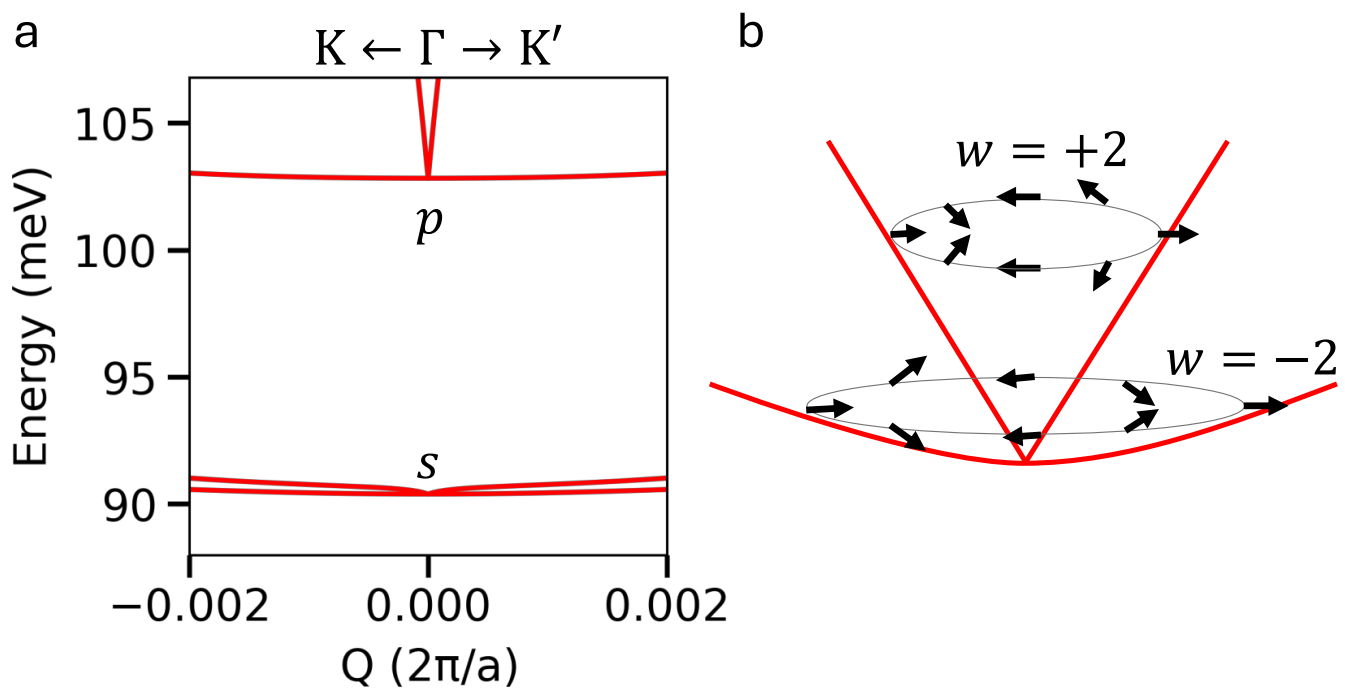

**Fig. 4.** a) Exciton band dispersion near $\mathbf{Q} = 0$ along the K-Γ-K′ path. The valley-degenerate doublet at $\mathbf{Q} = 0$ splits at finite $\mathbf{Q}$ due to inter-valley exchange interactions. The $p$-exciton shows a nonanalytic (v-shaped) upper branch and a parabolic lower branch, while the $s$-exciton exhibits weaker splitting. b) Valley pseudospin textures of the $p$-exciton branches. The upper (lower) branch carries a pseudospin winding number of +2 (−2).

## Tables

**Table 1: hBN-encapsulated BBG with applied field of 0.10 eV/Å.** Values of the valley g-factor for $s$- and $p$-excitons obtained from (a) band-edge single-particle approximation, (b) envelope-function-weighted single-particle moments, (c) full quantum formalism – two band approximation, (d) full quantum formalism – multi-band, and (e) experimental measurements (*14*). The $g$-factors are dimensionless.

| | $g_s$ | $g_p$ |
|---|---|---|
| a) Free electron-hole pair approximation with no Berry-curvature effect (Eq. 6) | 15.06 | 15.06 |
| b) Exciton envelope function averaged over k-space (Eq. 7) | 18.45 | 13.03 |
| c) Full quantum mechanical treatment – two-band (sum of Eqs. 9, 10, 11 and Eq. 8) | 20.25 | 1.93 |
| d) Full quantum mechanical treatment – multi-band (Eq. 4 and Eq. 8) | 20.32 | 1.79 |
| e) Experimental measured values (14) | $19.8 \pm 0.1$ | $1.4 \pm 0.8$ |

**Table 2:** Contributions to $\boldsymbol{\mu}_{n\mathbf{Q}}^{X}$ at $\mathbf{Q} = 0$ for *s*- and *p*-excitons of BBG, resolved into single-particle ($\boldsymbol{\mu}_{n\mathbf{Q}}^{X,sp}$), relative-motion ($\boldsymbol{\mu}_{n\mathbf{Q}}^{X,rel}$), and center-of-mass ($\boldsymbol{\mu}_{n\mathbf{Q}}^{X,COM}$) terms (Eq. 5). The first two rows sum to $\boldsymbol{\mu}_{n\mathbf{Q}}^{X,sp}$. The final row gives the total orbital magnetic moment. Both the *s* and *p* excitons at zero field are two-fold degenerate with moments of opposite sign; only the moment of the K-valley exciton is shown. All values are in $\mu_B$.

| | $s$ | $p$ |
|---|---|---|
| Envelope function average | 9.23 | 6.52 |
| Berry phase correction to e and h moments | 2.23 | 3.32 |
| e-h relative moment | −0.08 | −5.95 |
| e-h COM moment | −1.22 | −4.78 |
| Total exciton orbital magnetic moment | 10.16 | −0.89 |

# Supplementary Materials for

## Quantum Theory of Exciton Magnetic Moment: Interaction and Topological Effects

Gurjyot Sethi[1,2], Jiawei Ruan[1,2], Fang Zhang[1,2], Weichen Tang[1,2], Chen Hu[1,2], Mit Naik[1,2], and Steven G. Louie[1,2*]

[1]Department of Physics, University of California at Berkeley, Berkeley, CA, USA.
[2]Materials Sciences Division, Lawrence Berkeley National Laboratory, Berkeley, CA, USA.
*Corresponding author: sglouie@berkeley.edu (S.G.L.)

**Contents:**

## 1. First-order perturbation to quasi-particle Hamiltonian

We examine the first-order perturbation to the quasi-particle Hamiltonian $H$ describing a band electron in a periodic potential such as that of the Dyson's equation in the GW approach (1) in the quasiparticle approximation. When a uniform external magnetic field $\mathbf{B}$ is applied, the perturbed Hamiltonian to first order takes the form:

$$\widetilde{H} = H + \frac{e}{2}(\hat{\mathbf{v}} \cdot \mathbf{A} + \mathbf{A} \cdot \hat{\mathbf{v}}), \tag{S1}$$

where $\mathbf{A}$ is the vector potential. Choosing the symmetric gauge, $\mathbf{A} = \frac{1}{2}(\mathbf{B} \times \mathbf{r})$, the Hamiltonian simplifies to:

$$\widetilde{H} = H - \hat{\boldsymbol{\mu}} \cdot \mathbf{B}, \tag{S2}$$

where $\hat{\boldsymbol{\mu}} = -\frac{e}{2}\hat{\mathbf{r}} \times \hat{\mathbf{v}}$ is the orbital part of the magnetic moment operator. The quasiparticle eigenstates of $H$ are the Bloch wavefunctions, $|n\mathbf{k}\rangle = e^{i\mathbf{k}\cdot\hat{\mathbf{r}}}\,|u_{n\mathbf{k}}\rangle$ with the coordinate representation of the Bloch states given by $\langle \mathbf{r}|n\mathbf{k}\rangle = \langle \mathbf{r}|e^{i\mathbf{k}\cdot\hat{\mathbf{r}}}|u_{n\mathbf{k}}\rangle = e^{i\mathbf{k}\cdot\mathbf{r}}\langle \mathbf{r}|u_{n\mathbf{k}}\rangle = e^{i\mathbf{k}\cdot\mathbf{r}}u_{n\mathbf{k}}(\mathbf{r})$. The corresponding band energies are $\epsilon_{n\mathbf{k}}$. Using first-order perturbation theory, the perturbed quasiparticle energies are given by

$$\tilde{\epsilon}_{n\mathbf{k}} = \langle n\mathbf{k}|\widetilde{H}|n\mathbf{k}\rangle = \epsilon_{n\mathbf{k}} - \langle n\mathbf{k}|\hat{\boldsymbol{\mu}}|n\mathbf{k}\rangle \cdot \mathbf{B} = \epsilon_{n\mathbf{k}} - \boldsymbol{\mu}_{n\mathbf{k}} \cdot \mathbf{B}, \tag{S3}$$

where $\boldsymbol{\mu}_{n\mathbf{k}}$ is the single-particle orbital magnetic moment associated with state $|n\mathbf{k}\rangle$. The perturbed quasiparticle wavefunctions are given by

$$|\widetilde{n\mathbf{k}}\rangle = |n\mathbf{k}\rangle - \sum_{n'\mathbf{k}' \neq n\mathbf{k}} \frac{\langle n'\mathbf{k}'|\hat{\boldsymbol{\mu}}|n\mathbf{k}\rangle \cdot \mathbf{B}}{\epsilon_{n\mathbf{k}} - \epsilon_{n'\mathbf{k}'}} |n'\mathbf{k}'\rangle. \tag{S4}$$

## 2. Perturbed excitonic Hamiltonian in unperturbed excitonic basis

In the present work, for simplicity, we explicitly drop the spin Zeeman contribution to the magnetic moment, as spin-dependent terms vanish for the like-spin-transitions excitons (optically bright) considered here. But the spin contributions can be straightforwardly integrated into our formalism if required. We start with unperturbed excitonic Hamiltonian expressed in the free e-h band state basis, $|vc\mathbf{k}\mathbf{Q}\rangle = \left|c\mathbf{k} + \frac{\mathbf{Q}}{2}\right\rangle \otimes \left|v\mathbf{k} - \frac{\mathbf{Q}}{2}\right\rangle^*$, as described in Eq. 1, with $v$ and $c$ denote the valence and conduction bands, respectively. The completeness relationship for this basis is given by

$$\sum_{vc\mathbf{k}} |vc\mathbf{k}\mathbf{Q}\rangle\langle vc\mathbf{k}\mathbf{Q}| = 1. \tag{S5}$$

Under a magnetic field, the valence and conduction band wavefunctions are perturbed as described by Eq. S4, which would change the free e-h basis, now given by $|\widetilde{vc\mathbf{kQ}}\rangle = |\widetilde{c\mathbf{k}+\mathbf{Q}/2}\rangle \otimes |\widetilde{v\mathbf{k}-\mathbf{Q}/2}\rangle^*$, also forming a complete basis:

$$\sum_{vc\mathbf{kQ}} |\widetilde{vc\mathbf{kQ}}\rangle\langle\widetilde{vc\mathbf{kQ}}| = 1. \tag{S6}$$

The corresponding perturbed excitonic Hamiltonian in this perturbed basis is hence expressed within the GW-BSE approach as

$$\widetilde{H}^X_{\widetilde{vc\mathbf{kQ}},\widetilde{v'c'\mathbf{k}'\mathbf{Q}'}} = \langle\widetilde{vc\mathbf{kQ}}|\widetilde{H}^X|\widetilde{v'c'\mathbf{k}'\mathbf{Q}'}\rangle = \left(\tilde{\epsilon}_{\widetilde{c\mathbf{k}+\mathbf{Q}/2}} - \tilde{\epsilon}_{\widetilde{v\mathbf{k}-\mathbf{Q}/2}}\right)\delta_{\widetilde{vv'}}\delta_{\widetilde{cc'}}\delta_{\widetilde{\mathbf{kk}'}}\delta_{\widetilde{\mathbf{QQ}'}} + \widetilde{K}^{eh}_{\widetilde{vc\mathbf{kQ}},\widetilde{v'c'\mathbf{k}'\mathbf{Q}'}}. \tag{S7}$$

We note that for a general magnetic field perturbation, $\widetilde{H}^X$ is no longer diagonal in $\mathbf{Q}$. To evaluate $\langle X_{n\mathbf{Q}}|\widetilde{H}^X|X_{n'\mathbf{Q}'}\rangle$ between the unperturbed exciton states, we first use the completeness relation of Eq. S6:

$$\langle X_{n\mathbf{Q}}|\widetilde{H}^X|X_{n'\mathbf{Q}'}\rangle = \sum_{\widetilde{vc\mathbf{kQ}},\widetilde{v'c'\mathbf{k}'\mathbf{Q}'}} \langle X_{n\mathbf{Q}}|\widetilde{vc\mathbf{kQ}}\rangle\langle\widetilde{vc\mathbf{kQ}}|\widetilde{H}^X|\widetilde{v'c'\mathbf{k}'\mathbf{Q}'}\rangle\langle\widetilde{v'c'\mathbf{k}'\mathbf{Q}'}|X_{n'\mathbf{Q}'}\rangle. \tag{S8}$$

Next, we use the completeness relation of Eq. S5:

$$\langle X_{n\mathbf{Q}}|\widetilde{H}^X|X_{n'\mathbf{Q}'}\rangle = \sum_{\substack{vc\mathbf{k},\\ v'c'\mathbf{k}'}} \sum_{\substack{\widetilde{vc\mathbf{kQ}},\\ \widetilde{v'c'\mathbf{k}'\mathbf{Q}'}}} \langle X_{n\mathbf{Q}}|vc\mathbf{kQ}\rangle\langle vc\mathbf{kQ}|\widetilde{vc\mathbf{kQ}}\rangle\langle\widetilde{vc\mathbf{kQ}}|\widetilde{H}^X|\widetilde{v'c'\mathbf{k}'\mathbf{Q}'}\rangle\langle\widetilde{v'c'\mathbf{k}'\mathbf{Q}'}|v'c'\mathbf{k}'\mathbf{Q}'\rangle\langle v'c'\mathbf{k}'\mathbf{Q}'|X_{n'\mathbf{Q}'}\rangle \tag{S9}$$

Using Eq. 2 of the main text,

$$\langle X_{n\mathbf{Q}}|\widetilde{H}^X|X_{n'\mathbf{Q}'}\rangle = \sum_{\substack{vc\mathbf{k},\\ v'c'\mathbf{k}'}} A^{n\mathbf{Q}*}_{vc\mathbf{k}} A^{n\mathbf{Q}'}_{v'c'\mathbf{k}'} [\sum_{\substack{\widetilde{vc\mathbf{kQ}},\\ \widetilde{v'c'\mathbf{k}'\mathbf{Q}'}}} \langle vc\mathbf{kQ}|\widetilde{vc\mathbf{kQ}}\rangle\langle\widetilde{vc\mathbf{kQ}}|\widetilde{H}^X|\widetilde{v'c'\mathbf{k}'\mathbf{Q}'}\rangle\langle\widetilde{v'c'\mathbf{k}'\mathbf{Q}'}|v'c'\mathbf{k}'\mathbf{Q}'\rangle]. \tag{S10}$$

The expression within the square backets in the above equation can be expanded into two terms using Eq. S7,

$$\sum_{\widetilde{vc\mathbf{kQ}},\widetilde{v'c'\mathbf{k}'\mathbf{Q}'}} \langle vc\mathbf{kQ}|\widetilde{vc\mathbf{kQ}}\rangle \left\langle \widetilde{vc\mathbf{kQ}}\middle|\widetilde{H}^X\middle|\widetilde{v'c'\mathbf{k}'\mathbf{Q}'}\right\rangle \left\langle \widetilde{v'c'\mathbf{k}'\mathbf{Q}'}\middle|v'c'\mathbf{k}'\mathbf{Q}'\right\rangle =$$

$$\sum_{\widetilde{vc\mathbf{kQ}},\widetilde{v'c'\mathbf{k}'\mathbf{Q}'}} \langle vc\mathbf{kQ}|\widetilde{vc\mathbf{kQ}}\rangle\{(\tilde{\epsilon}_{c\widetilde{\mathbf{k}+\mathbf{Q}/2}} - \tilde{\epsilon}_{v\widetilde{\mathbf{k}-\mathbf{Q}/2}})\delta_{\widetilde{vv'}}\delta_{\widetilde{cc'}}\delta_{\widetilde{\mathbf{kk}'}}\delta_{\widetilde{\mathbf{QQ}'}} + \widetilde{K}^{eh}_{\widetilde{vc\mathbf{kQ}},\widetilde{v'c'\mathbf{k}'\mathbf{Q}'}}\} \left\langle \widetilde{v'c'\mathbf{k}'\mathbf{Q}'}\middle|v'c'\mathbf{k}'\mathbf{Q}'\right\rangle =$$

$$\sum_{\widetilde{vc\mathbf{kQ}}} \langle vc\mathbf{kQ}|\widetilde{vc\mathbf{kQ}}\rangle(\tilde{\epsilon}_{c\widetilde{\mathbf{k}+\mathbf{Q}/2}} - \tilde{\epsilon}_{v\widetilde{\mathbf{k}-\mathbf{Q}/2}})\langle\widetilde{vc\mathbf{kQ}}|v'c'\mathbf{k}'\mathbf{Q}'\rangle + \tag{S11}$$

$$+ \sum_{\widetilde{vc\mathbf{kQ}},\widetilde{v'c'\mathbf{k}'\mathbf{Q}'}} \langle vc\mathbf{kQ}|\widetilde{vc\mathbf{kQ}}\rangle \widetilde{K}^{eh}_{\widetilde{vc\mathbf{kQ}},\widetilde{v'c'\mathbf{k}'\mathbf{Q}'}} \left\langle \widetilde{v'c'\mathbf{k}'\mathbf{Q}'}\middle|v'c'\mathbf{k}'\mathbf{Q}'\right\rangle$$

Focusing on the second term (on the RHS of the second equal sign) in the above expression, we expand $\widetilde{K}^{eh}_{\widetilde{vc\mathbf{kQ}},\widetilde{v'c'\mathbf{k'}\mathbf{Q'}}} = \langle\widetilde{vc\mathbf{kQ}}|\widetilde{K}^{eh}|\widetilde{v'c'\mathbf{k'}\mathbf{Q'}}\rangle$. Note that the perturbed e-h kernel matrix elements involves the perturbed quasiparticle wavefunctions as well as the perturbed screening of the e-h interaction. We further assume that the screening of the Coulomb interaction between electron and hole does not change under a weak **B** field, i.e., $W \approx \widetilde{W}$ or equivalently $\widetilde{K}^{eh} \approx K^{eh}$. We have for this term

$$\sum_{\widetilde{vc\mathbf{kQ}},\widetilde{v'c'\mathbf{k}'\mathbf{Q}'}} \langle vc\mathbf{kQ}|\widetilde{vc\mathbf{kQ}}\rangle \widetilde{K}^{eh}_{\widetilde{vc\mathbf{kQ}},\widetilde{v'c'\mathbf{k}'\mathbf{Q}'}} \left\langle \widetilde{v'c'\mathbf{k}'\mathbf{Q}'}\middle|v'c'\mathbf{k}'\mathbf{Q}'\right\rangle =$$

$$\sum_{\widetilde{vc\mathbf{kQ}},\widetilde{v'c'\mathbf{k}'\mathbf{Q}'}} \langle vc\mathbf{kQ}|\widetilde{vc\mathbf{kQ}}\rangle \left\langle \widetilde{vc\mathbf{kQ}}\middle|\widetilde{K}^{eh}\middle|\widetilde{v'c'\mathbf{k}'\mathbf{Q}'}\right\rangle \left\langle \widetilde{v'c'\mathbf{k}'\mathbf{Q}'}\middle|v'c'\mathbf{k}'\mathbf{Q}'\right\rangle \approx \tag{S12}$$

$$\sum_{\widetilde{vc\mathbf{kQ}},\widetilde{v'c'\mathbf{k}'\mathbf{Q}'}} \langle vc\mathbf{kQ}|\widetilde{vc\mathbf{kQ}}\rangle \left\langle \widetilde{vc\mathbf{kQ}}\middle|K^{eh}\middle|\widetilde{v'c'\mathbf{k}'\mathbf{Q}'}\right\rangle \left\langle \widetilde{v'c'\mathbf{k}'\mathbf{Q}'}\middle|v'c'\mathbf{k}'\mathbf{Q}'\right\rangle =$$

$$\left\langle vc\mathbf{kQ}\middle|K^{eh}\middle|v'c'\mathbf{k}'\mathbf{Q}'\right\rangle = \delta_{\mathbf{QQ}'}K^{eh}_{vc\mathbf{k},v'c'\mathbf{k}'}(\mathbf{Q})$$

In the last line we have used the fact that the unperturbed kernel is diagonal in **Q**. Next, for the first term on the RHS after the second equal sign in Eq. S11, we need to evaluate $\langle vc\mathbf{kQ}|\widetilde{vc\mathbf{kQ}}\rangle$ which is $\langle vc\mathbf{kQ}|\widetilde{vc\mathbf{kQ}}\rangle = \langle v\widetilde{\mathbf{k}-\mathbf{Q}}/2\,|v\mathbf{k}-\mathbf{Q}/2\rangle\langle c\mathbf{k}+\mathbf{Q}/2\,|c\widetilde{\mathbf{k}+\mathbf{Q}}/2\rangle$, i.e., we need the overlaps between unperturbed and perturbed single-particle wavefunctions which can be computed using Eq. S4. Following this, after straightforward algebraic manipulations we obtain Eq. 3 in the main text.

## 3. Position operator in Bloch band state basis

To illustrate the divergent behavior of the position operator $\hat{\mathbf{r}}$ in Bloch band state basis, we explicitly evaluate the matrix elements $\langle n'\mathbf{k}'|\hat{\mathbf{r}}|n\mathbf{k}\rangle$. We begin by differentiating the Bloch state with respect to **k**:

$$\nabla_{\mathbf{k}}|n\mathbf{k}\rangle = \nabla_{\mathbf{k}}(e^{i\mathbf{k}\cdot\hat{\mathbf{r}}}|u_{n\mathbf{k}}\rangle) = i\hat{\mathbf{r}}|n\mathbf{k}\rangle + e^{i\mathbf{k}\cdot\hat{\mathbf{r}}}|\nabla_{\mathbf{k}}u_{n\mathbf{k}}\rangle. \tag{S13}$$

Solving for $\hat{\mathbf{r}}|n\mathbf{k}\rangle$ and substituting into the matrix element yields:

$$\begin{aligned}\langle n'\mathbf{k}'|\hat{\mathbf{r}}|n\mathbf{k}\rangle &= \langle n'\mathbf{k}'|\big(-i\nabla_{\mathbf{k}}|n\mathbf{k}\rangle + ie^{i\mathbf{k}\cdot\hat{\mathbf{r}}}|\nabla_{\mathbf{k}}u_{n\mathbf{k}}\rangle\big)\\ &= -i\nabla_{\mathbf{k}}(\langle n'\mathbf{k}'|n\mathbf{k}\rangle) + i\langle u_{n'\mathbf{k}'}|e^{i(\mathbf{k}-\mathbf{k}')\cdot\hat{\mathbf{r}}}|\nabla_{\mathbf{k}}u_{n\mathbf{k}}\rangle\\ &= -i\delta_{nn'}(\nabla_{\mathbf{k}}\delta_{\mathbf{k}\mathbf{k}'}) + i\big\langle u_{n'\mathbf{k}'}\big|e^{i(\mathbf{k}-\mathbf{k}')\cdot\hat{\mathbf{r}}}\big|\nabla_{\mathbf{k}}u_{n\mathbf{k}}\big\rangle.\end{aligned} \tag{S14}$$

The second term involves an overlap integral over all space. Since $u_{n\mathbf{k}}(\mathbf{r})$ is a periodic function in $\mathbf{r}$, we partition real space into unit cells $\mathbf{r_{uc}}$, and write $\tilde{\mathbf{r}} = \mathbf{r} - \mathbf{r_{uc}}$, giving:

$$\begin{aligned}\big\langle u_{n'\mathbf{k}'}\big|e^{i(\mathbf{k}-\mathbf{k}')\cdot\hat{\mathbf{r}}}\big|\nabla_{\mathbf{k}}u_{n\mathbf{k}}\big\rangle &= \int d\mathbf{r}\, u^*_{n'\mathbf{k}'}(\mathbf{r})e^{i(\mathbf{k}-\mathbf{k}')\cdot\mathbf{r}}\nabla_{\mathbf{k}}u_{n\mathbf{k}}(\mathbf{r})\\ &= \sum_{\mathbf{r_{uc}}} e^{i(\mathbf{k}-\mathbf{k}')\cdot\mathbf{r_{uc}}}\int d\mathbf{r}\, u^*_{n'\mathbf{k}'}(\mathbf{r})e^{i(\mathbf{k}-\mathbf{k}')\cdot\tilde{\mathbf{r}}}\nabla_{\mathbf{k}}u_{n\mathbf{k}}(\mathbf{r})\\ &= \sum_{\mathbf{r_{uc}}} e^{i(\mathbf{k}-\mathbf{k}')\cdot\mathbf{r_{uc}}}\int_{uc} d\tilde{\mathbf{r}}\, u^*_{n'\mathbf{k}'}(\tilde{\mathbf{r}})e^{i(\mathbf{k}-\mathbf{k}')\cdot\tilde{\mathbf{r}}}\nabla_{\mathbf{k}}u_{n\mathbf{k}}(\tilde{\mathbf{r}})\\ &= \delta_{\mathbf{k}\mathbf{k}'}\langle u_{n'\mathbf{k}}|\nabla_{\mathbf{k}}u_{n\mathbf{k}}\rangle_{uc},\end{aligned} \tag{S15}$$

where integrals go only within a unit-cell and remain the same for every unit-cell due to the periodicity of the cell-periodic functions. We have made use of the identity $\sum_{\mathbf{r_{uc}}} e^{i(\mathbf{k}-\mathbf{k}')\cdot\mathbf{r_{uc}}} = \delta_{\mathbf{k}\mathbf{k}'}$.

Combining these results, the full matrix element becomes:

$$\langle n'\mathbf{k}'|\hat{\mathbf{r}}|n\mathbf{k}\rangle = -i\delta_{nn'}(\nabla_{\mathbf{k}}\delta_{\mathbf{k}\mathbf{k}'}) + i\delta_{\mathbf{k}\mathbf{k}'}\langle u_{n'\mathbf{k}}|\nabla_{\mathbf{k}}u_{n\mathbf{k}}\rangle_{uc}. \tag{S16}$$

This expression reveals two key issues in the intra-band ($n = n'$) channel: (i) the first term $-i\nabla_{\mathbf{k}}\delta(\boldsymbol{k} - k')$ is explicitly divergent, and (ii) the second term which is the Berry connection term is gauge dependent. These issues arise fundamentally from the unbounded nature of $\hat{\mathbf{r}}$ and the extended nature of Bloch wavefunctions underpinning the need for a regularization scheme such as the periodic analog $\hat{\mathbf{r}} \to \hat{\mathbf{r}}_{\boldsymbol{\lambda}}{}' = -i\nabla_{\boldsymbol{\lambda}}e^{i\boldsymbol{\lambda}\cdot\hat{\mathbf{r}}}$ discussed in the main text. Using this substitution, the excitonic matrix element in Eq. 3 can be rewritten rigorously as:

$$\begin{aligned}\big\langle X_{n\mathbf{Q}}\big|\widetilde{H}^X\big|X_{n'\mathbf{Q}'}\big\rangle &= \delta_{\mathbf{Q},\mathbf{Q}'}\epsilon^X_{n\mathbf{Q}} +\\ \lim_{\boldsymbol{\lambda}\to 0}\frac{e}{2i}\mathbf{B}\cdot\Big(\nabla_{\boldsymbol{\lambda}}\times\sum_{vc\mathbf{k}}\Big[&\sum_{c'\mathbf{k}'} A^{n\mathbf{Q}*}_{vc\mathbf{k}}A^{n'\mathbf{Q}'}_{vc'\mathbf{k}'}\delta_{\mathbf{k}-\frac{\mathbf{Q}}{2},\mathbf{k}'-\frac{Q'}{2}}\Big\langle c\mathbf{k}+\frac{\mathbf{Q}}{2}\Big|\,e^{i\boldsymbol{\lambda}\cdot\hat{\mathbf{r}}}\hat{\mathbf{v}}\,\Big|c'\mathbf{k}'+\frac{\mathbf{Q}'}{2}\Big\rangle\\ &-\sum_{v'\mathbf{k}'} A^{n\mathbf{Q}*}_{vc\mathbf{k}}A^{n'\mathbf{Q}'}_{v'c\mathbf{k}'}\delta_{\mathbf{k}+\frac{\mathbf{Q}}{2},\mathbf{k}'+\frac{Q'}{2}}\Big\langle v'\mathbf{k}'-\frac{\mathbf{Q}'}{2}\Big|\,e^{i\boldsymbol{\lambda}\cdot\hat{\mathbf{r}}}\hat{\mathbf{v}}\,\Big|v\mathbf{k}-\frac{\mathbf{Q}}{2}\Big\rangle\Big]\Big)\end{aligned} \tag{S17}$$

Evaluating the matrix elements yields $\big\langle n\mathbf{k}\big|e^{i\boldsymbol{\lambda}\cdot\hat{\mathbf{r}}}\hat{\mathbf{v}}\big|n'\mathbf{k}'\big\rangle = \delta_{\mathbf{k}',\mathbf{k}-\boldsymbol{\lambda}}\langle n\mathbf{k}|\hat{\mathbf{v}}(\mathbf{k}-\boldsymbol{\lambda})|n'\mathbf{k}-\boldsymbol{\lambda}\rangle_{uc}$, with $\hat{\mathbf{v}}(\mathbf{k}) = e^{-i\mathbf{k}\cdot\hat{\mathbf{r}}}\hat{\mathbf{v}}e^{i\mathbf{k}\cdot\hat{\mathbf{r}}}$. This will lead to quasi momentum conservation rules that relate $\mathbf{Q}$ and $\mathbf{Q}'$, as well as $\mathbf{k}$ and $\mathbf{k}'$. In the following we drop the subscript *uc* on the overlaps for readability. Using differentiation by parts afterwards one can derive:

$$\langle X_{n\mathbf{Q}}|\tilde{H}^X|X_{n'\mathbf{Q}'}\rangle = \delta_{\mathbf{Q},\mathbf{Q}'}E^X_{n\mathbf{Q}} - \delta_{\mathbf{Q},\mathbf{Q}'}\mathbf{B}\cdot\sum_{vc\mathbf{k}}\left\{\sum_{c'} A^{n\mathbf{Q}*}_{vc\mathbf{k}}A^{n'\mathbf{Q}}_{vc'\mathbf{k}}\boldsymbol{\mu}_{cc'\mathbf{k}+\frac{\mathbf{Q}}{2}} - \sum_{v'} A^{n\mathbf{Q}*}_{vc\mathbf{k}}A^{n'\mathbf{Q}}_{v'c\mathbf{k}}\boldsymbol{\mu}'_{v'v\mathbf{k}-\frac{\mathbf{Q}}{2}}\right\}$$

$$-\frac{e}{2\hbar}\delta_{\mathbf{Q},\mathbf{Q}'}\mathbf{B}\cdot\sum_{vc\mathbf{k}}\left\{\sum_{c'} A^{n\mathbf{Q}*}_{vc\mathbf{k}}A^{n'\mathbf{Q}}_{vc'\mathbf{k}}\left(\nabla_{\mathbf{k}}\epsilon_{c\mathbf{k}}|_{\mathbf{k}+\frac{\mathbf{Q}}{2}}\times\boldsymbol{\mathcal{A}}_{cc'\mathbf{k}+\frac{\mathbf{Q}}{2}}\right) - \sum_{v'} A^{n\mathbf{Q}*}_{vc\mathbf{k}}A^{n'\mathbf{Q}}_{v'c\mathbf{k}}\left(\nabla_{\mathbf{k}}\epsilon_{v\mathbf{k}}|_{\mathbf{k}-\frac{\mathbf{Q}}{2}}\times\boldsymbol{\mathcal{A}}_{v'v\mathbf{k}-\frac{\mathbf{Q}}{2}}\right)\right\}$$

$$-\frac{e}{4i}\delta_{\mathbf{Q},\mathbf{Q}'}\mathbf{B}\cdot\sum_{vc\mathbf{k}}\left\{\sum_{c'}\left(A^{n\mathbf{Q}*}_{vc\mathbf{k}}\nabla_{\mathbf{k}}A^{n'\mathbf{Q}}_{vc'\mathbf{k}}\times\mathbf{v}_{cc'\mathbf{k}+\frac{\mathbf{Q}}{2}}\right) + \sum_{v'}\left(A^{n\mathbf{Q}*}_{vc\mathbf{k}}\nabla_{\mathbf{k}}A^{n'\mathbf{Q}}_{v'c\mathbf{k}}\times\mathbf{v}_{v'v\mathbf{k}-\frac{\mathbf{Q}}{2}}\right)\right\} \quad \text{(S18)}$$

$$-\frac{e}{2i}\delta_{\mathbf{Q},\mathbf{Q}'}\mathbf{B}\cdot\sum_{vc\mathbf{k}}\left\{\sum_{c'}\left(A^{n\mathbf{Q}*}_{vc\mathbf{k}}\nabla_{\mathbf{Q}}A^{n'\mathbf{Q}}_{vc'\mathbf{k}}\times\mathbf{v}_{cc'\mathbf{k}+\frac{\mathbf{Q}}{2}}\right) - \sum_{v'}\left(A^{n\mathbf{Q}*}_{vc\mathbf{k}}\nabla_{\mathbf{Q}}A^{n'\mathbf{Q}}_{v'c\mathbf{k}}\times\mathbf{v}_{v'v\mathbf{k}-\frac{\mathbf{Q}}{2}}\right)\right\}$$

$$-\frac{e}{2i}\mathbf{B}\cdot\left((\nabla_{\mathbf{Q}}\delta_{\mathbf{Q},\mathbf{Q}'})\times\sum_{vc\mathbf{k}}\left\{\sum_{c'} A^{n\mathbf{Q}*}_{vc\mathbf{k}}A^{n'\mathbf{Q}}_{vc'\mathbf{k}}\mathbf{v}_{cc'\mathbf{k}+\frac{\mathbf{Q}}{2}} - \sum_{v'} A^{n\mathbf{Q}*}_{vc\mathbf{k}}A^{n'\mathbf{Q}}_{v'c\mathbf{k}}\mathbf{v}_{v'v\mathbf{k}-\frac{\mathbf{Q}}{2}}\right\}\right),$$

with same notations as in Eq. 4.

## 4. Perturbed Hamiltonian in exciton wave-packet basis

To resolve the divergence associated with the perturbed exciton Hamiltonian in the exciton Bloch basis in Eq. S18, we construct a localized exciton wave packet centered at $\mathbf{R_c}$ in the real-space COM coordinates. This wave packet is defined as a superposition of exciton Bloch states within a band $n$:

$$|W_{n\mathbf{Q_c}}\rangle = \sum_{\mathbf{Q}} w(\mathbf{Q}-\mathbf{Q_c})|X_{n\mathbf{Q}}\rangle, \quad \text{(S19)}$$

where the weighting function $w(\mathbf{Q}-\mathbf{Q_c})$ is localized around a central COM momentum $\mathbf{Q_c}$ such that:

$$\sum_{\mathbf{Q}} \mathbf{Q}|w(\mathbf{Q}-\mathbf{Q_c})|^2 = \mathbf{Q_c}, \quad \text{(S20)}$$

and the wave packet is centered in real space at:

$$\langle W_{n\mathbf{Q_c}}|\mathbf{R}|W_{n\mathbf{Q_c}}\rangle = \mathbf{R_c}. \quad \text{(S21)}$$

We note here that we create the wave-packet for a single exciton band. One can create a coherent wave-packet for multiple bands (*2*) and follow the steps shown below to arrive at an exciton orbital magnetic moment matrix in the basis of the multiple bands considered; diagonalization of which will result in the orbital magnetic moments of exciton states in each of the bands. For the physical picture considered in this work, i.e., the valley Zeeman splitting of $\mathbf{Q}=0$ excitons, our results for exciton wave-packet constructed using two exciton bands agree with those obtained using a single exciton band. Hence, we only illustrate the latter here.

Expanding the expectation value in Eq. S21 gives:

$$\langle W_{n\mathbf{Q_c}}|\mathbf{R}|W_{n\mathbf{Q_c}}\rangle = \sum_{\mathbf{Q}}\sum_{\mathbf{Q}'} w^*(\mathbf{Q}-\mathbf{Q_c})w(\mathbf{Q}'-\mathbf{Q_c})\langle X_{n\mathbf{Q}}|\hat{\mathbf{R}}|X_{n\mathbf{Q}'}\rangle. \tag{S22}$$

Given the exciton Bloch form $|X_{n\mathbf{Q}}\rangle = e^{i\mathbf{Q}.\hat{\mathbf{R}}}|u^X_{n\mathbf{Q}}\rangle$, we may use the equivalence of the result in Eq. S16 to evaluate:

$$\begin{aligned}
\langle W_{n\mathbf{Q_c}}|\mathbf{R}|W_{n\mathbf{Q_c}}\rangle &= \sum_{\mathbf{Q}}\sum_{\mathbf{Q}'} w^*(\mathbf{Q}-\mathbf{Q_c})w(\mathbf{Q}'-\mathbf{Q_c})[-i\nabla_{\mathbf{Q}'}\delta_{\mathbf{QQ}'} + i\delta_{\mathbf{QQ}'}\langle u^X_{n\mathbf{Q}}|\nabla_{\mathbf{Q}}u^X_{n\mathbf{Q}}\rangle] \\
&= i\sum_{\mathbf{Q}}\sum_{\mathbf{Q}'} w^*(\mathbf{Q}-\mathbf{Q_c})w(\mathbf{Q}'-\mathbf{Q_c})\nabla_{\mathbf{Q}}\delta_{\mathbf{QQ}'} + i\sum_{\mathbf{Q}}\sum_{\mathbf{Q}'} w^*(\mathbf{Q}-\mathbf{Q_c})w(\mathbf{Q}'-\mathbf{Q_c})\delta_{\mathbf{QQ}'}\langle u^X_{n\mathbf{Q}}|\nabla_{\mathbf{Q}}u^X_{n\mathbf{Q}}\rangle \\
&= i\sum_{\mathbf{Q}} w^*(\mathbf{Q}-\mathbf{Q_c})\nabla_{\mathbf{Q}}\left(\sum_{Q'} w(\mathbf{Q}')\delta_{\mathbf{QQ}'}\right) + i\sum_{\mathbf{Q}}\left|w(\mathbf{Q}-\mathbf{Q_c})\right|^2\langle u^X_{n\mathbf{Q}}|\nabla_{\mathbf{Q}}u^X_{n\mathbf{Q}}\rangle \\
&= i\sum_{\mathbf{Q}} w^*(\mathbf{Q}-\mathbf{Q_c})\nabla_{\mathbf{Q}}w(\mathbf{Q}) + \sum_{\mathbf{Q}}\left|w(\mathbf{Q}-\mathbf{Q_c})\right|^2\boldsymbol{\mathcal{A}}^{\mathrm{X}}_{n\mathbf{Q}},
\end{aligned} \tag{S23}$$

where $\boldsymbol{\mathcal{A}}^{\mathrm{X}}_{n\mathbf{Q}} = i\langle u^X_{n\mathbf{Q}}|\nabla_{\mathbf{Q}}u^X_{n\mathbf{Q}}\rangle$ is the exciton band Berry connection. In terms of the **k**-space exciton envelope functions $A^{n\mathbf{Q}}_{vc\mathbf{k}}$ and single-particle band Berry connections $\boldsymbol{\mathcal{A}}_{cc'\mathbf{k}+\frac{\mathbf{Q}}{2}}$ and $\boldsymbol{\mathcal{A}}_{v'v\mathbf{k}-\frac{\mathbf{Q}}{2}}$:

$$\boldsymbol{\mathcal{A}}^{\mathrm{X}}_{n\mathbf{Q}} = i\sum_{vc\mathbf{k}} A^{n\mathbf{Q}*}_{vc\mathbf{k}}\nabla_{\mathbf{Q}}A^{n\mathbf{Q}}_{vc\mathbf{k}} + \sum_{vck}\left\{\frac{1}{2}\sum_{c'} A^{n\mathbf{Q}*}_{vc\mathbf{k}}A^{n\mathbf{Q}}_{vc'\mathbf{k}}\boldsymbol{\mathcal{A}}_{cc'\mathbf{k}+\frac{\mathbf{Q}}{2}} + \frac{1}{2}\sum_{v'} A^{n\mathbf{Q}*}_{vc\mathbf{k}}A^{n\mathbf{Q}}_{v'c\mathbf{k}}\boldsymbol{\mathcal{A}}_{v'v\mathbf{k}-\frac{\mathbf{Q}}{2}}\right\}. \tag{S24}$$

To analyze the response of the wave packet under a magnetic field, we adopt a vector potential $\mathbf{A}(\mathbf{r}) = \frac{1}{2}[\mathbf{B}\times(\mathbf{r}-\mathbf{R_c})]$ chosen such that $\mathbf{A}(\mathbf{R_c}) = 0$, enabling the use of first-order perturbation theory. The perturbed single-particle Hamiltonian becomes:

$$\tilde{H} = H + \frac{e}{2}\mathbf{B}\cdot[(\hat{\mathbf{r}}-\mathbf{R_c})\times\hat{\mathbf{v}}]. \tag{S25}$$

Correspondingly, the matrix elements of the perturbed excitonic Hamiltonian in the unperturbed exciton basis take the form:

$$\begin{aligned}
&\langle X_{n\mathbf{Q}}|\tilde{H}^X|X_{n'\mathbf{Q}'}\rangle = \delta_{\mathbf{Q},\mathbf{Q}'}\epsilon^X_{n\mathbf{Q}} - \\
&\mathbf{B}\cdot\sum_{vc\mathbf{k}}\left[\sum_{c'\mathbf{k}'} A^{n\mathbf{Q}*}_{vc\mathbf{k}}A^{n'\mathbf{Q}'}_{vc'\mathbf{k}'}\delta_{\mathbf{k}-\frac{\mathbf{Q}}{2},\mathbf{k}'-\frac{\mathbf{Q}'}{2}}\left\langle c\mathbf{k}+\frac{\mathbf{Q}}{2}|\hat{\boldsymbol{\mu}}|c'\mathbf{k}'+\frac{\mathbf{Q}'}{2}\right\rangle - \sum_{v'\mathbf{k}'} A^{n\mathbf{Q}*}_{vc\mathbf{k}}A^{n'\mathbf{Q}'}_{v'c\mathbf{k}'}\delta_{\mathbf{k}+\frac{\mathbf{Q}}{2},\mathbf{k}'+\frac{\mathbf{Q}'}{2}}\left\langle v'\mathbf{k}'-\frac{\mathbf{Q}'}{2}|\hat{\boldsymbol{\mu}}|v\mathbf{k}-\frac{\mathbf{Q}}{2}\right\rangle\right] \\
&-\frac{e}{2}\mathbf{B}\cdot\left(\mathbf{R_c}\times\sum_{vc\mathbf{k}}\left[\sum_{c'\mathbf{k}'} A^{n\mathbf{Q}*}_{vc\mathbf{k}}A^{n'\mathbf{Q}'}_{vc'\mathbf{k}'}\delta_{\mathbf{k}-\frac{\mathbf{Q}}{2},\mathbf{k}'-\frac{\mathbf{Q}'}{2}}\left\langle c\mathbf{k}+\frac{\mathbf{Q}}{2}|\hat{\mathbf{v}}|c'\mathbf{k}'+\frac{\mathbf{Q}'}{2}\right\rangle - \sum_{v'\mathbf{k}'} A^{n\mathbf{Q}*}_{vc\mathbf{k}}A^{n'\mathbf{Q}'}_{v'c\mathbf{k}'}\delta_{\mathbf{k}+\frac{\mathbf{Q}}{2},\mathbf{k}'+\frac{\mathbf{Q}'}{2}}\left\langle v'\mathbf{k}'-\frac{\mathbf{Q}'}{2}|\hat{\mathbf{v}}|v\mathbf{k}-\frac{\mathbf{Q}}{2}\right\rangle\right]\right)
\end{aligned} \tag{S26}$$

Following Eq. S17, Eq. S18, and Eq. S26, we evaluate the expectation value of the perturbed Hamiltonian in this specific gauge and in the exciton wave-packet basis:

$$
\begin{aligned}
\langle W_{\mathbf{Q_c}}|\tilde{H}^X|W_{\mathbf{Q_c}}\rangle &= \sum_{\mathbf{Q}}\sum_{\mathbf{Q}'} w^*(\mathbf{Q}-\mathbf{Q_c})w(\mathbf{Q}'-\mathbf{Q_c})\langle X_{n\mathbf{Q}}|\tilde{H}^X|X_{n\mathbf{Q}'}\rangle = \sum_{\mathbf{Q}}\sum_{\mathbf{Q}'} w^*(\mathbf{Q}-\mathbf{Q_c})w(\mathbf{Q}'-\mathbf{Q_c})\delta_{\mathbf{Q},\mathbf{Q}'}E^X_{n\mathbf{Q}} \\
&-\mathbf{B}\cdot\sum_{\mathbf{Q}}\sum_{\mathbf{Q}'} w^*(\mathbf{Q}-\mathbf{Q_c})w(\mathbf{Q}'-\mathbf{Q_c})\delta_{\mathbf{Q},\mathbf{Q}'}\sum_{vc\mathbf{k}}\left\{\sum_{c'} A^{n\mathbf{Q}*}_{vc\mathbf{k}}A^{n\mathbf{Q}}_{vc'\mathbf{k}}\boldsymbol{\mu}_{cc'\mathbf{k}+\frac{\mathbf{Q}}{2}} - \sum_{v'} A^{n\mathbf{Q}*}_{vc\mathbf{k}}A^{n\mathbf{Q}}_{v'c\mathbf{k}}\boldsymbol{\mu}'_{v'v\mathbf{k}-\frac{\mathbf{Q}}{2}}\right\} \\
&-\frac{e}{2\hbar}\mathbf{B}\cdot\sum_{\mathbf{Q}}\sum_{\mathbf{Q}'} w^*(\mathbf{Q}-\mathbf{Q_c})w(\mathbf{Q}-\mathbf{Q_c}')\delta_{\mathbf{Q},\mathbf{Q}'}\sum_{vc\mathbf{k}}\left\{\sum_{c'} A^{n\mathbf{Q}*}_{vc\mathbf{k}}A^{n\mathbf{Q}}_{vc'\mathbf{k}}(\nabla_{\mathbf{k}}\epsilon_{c\mathbf{k}}\times\boldsymbol{\mathcal{A}}_{cc'})_{\mathbf{k}+\frac{\mathbf{Q}}{2}} - \sum_{v'} A^{n\mathbf{Q}*}_{vc\mathbf{k}}A^{n\mathbf{Q}}_{v'c\mathbf{k}}(\nabla_{\mathbf{k}}\epsilon_{v\mathbf{k}}\times\boldsymbol{\mathcal{A}}_{v'v})_{\mathbf{k}-\frac{\mathbf{Q}}{2}}\right\} \\
&-\frac{e}{4i}\mathbf{B}\cdot\sum_{\mathbf{Q}}\sum_{\mathbf{Q}'} w^*(\mathbf{Q}-\mathbf{Q_c})w(\mathbf{Q}'-\mathbf{Q_c})\,\delta_{\mathbf{Q},\mathbf{Q}'}\cdot\sum_{vc\mathbf{k}}\left\{\sum_{c'}\left(A^{n\mathbf{Q}*}_{vc\mathbf{k}}\nabla_{\mathbf{k}}A^{n\mathbf{Q}}_{vc'\mathbf{k}}\times\mathbf{v}_{cc'\mathbf{k}+\frac{\mathbf{Q}}{2}}\right) + \sum_{v'}\left(A^{n\mathbf{Q}*}_{vc\mathbf{k}}\nabla_{\mathbf{k}}A^{n\mathbf{Q}}_{v'c\mathbf{k}}\times\mathbf{v}_{v'v\mathbf{k}-\frac{\mathbf{Q}}{2}}\right)\right\} \\
&-\frac{e}{2i}\mathbf{B}\cdot\sum_{\mathbf{Q}}\sum_{\mathbf{Q}'} w^*(\mathbf{Q}-\mathbf{Q_c})w(\mathbf{Q}'-\mathbf{Q_c})\,\delta_{\mathbf{Q},\mathbf{Q}'}\sum_{vc\mathbf{k}}\left\{\sum_{c'}\left(A^{n\mathbf{Q}*}_{vc\mathbf{k}}\nabla_{\mathbf{Q}}A^{n\mathbf{Q}}_{vc'\mathbf{k}}\times\mathbf{v}_{cc'\mathbf{k}+\frac{\mathbf{Q}}{2}}\right) - \sum_{v'}\left(A^{n\mathbf{Q}*}_{vc\mathbf{k}}\nabla_{\mathbf{Q}}A^{n\mathbf{Q}}_{v'c\mathbf{k}}\times\mathbf{v}_{v'v\mathbf{k}-\frac{\mathbf{Q}}{2}}\right)\right\} \\
&-\frac{e}{2i}\mathbf{B}\cdot\sum_{\mathbf{Q}}\sum_{\mathbf{Q}'} w^*(\mathbf{Q}-\mathbf{Q_c})w(\mathbf{Q}'-\mathbf{Q_c})\left((\nabla_{\mathbf{Q}}\delta_{\mathbf{Q},\mathbf{Q}'})\times\sum_{vc\mathbf{k}}\left\{\sum_{c'} A^{n\mathbf{Q}*}_{vc\mathbf{k}}A^{n\mathbf{Q}}_{vc'\mathbf{k}}\mathbf{v}_{cc'\mathbf{k}+\frac{\mathbf{Q}}{2}} - \sum_{v'} A^{n\mathbf{Q}*}_{vc\mathbf{k}}A^{n\mathbf{Q}}_{v'c\mathbf{k}}\mathbf{v}_{v'v\mathbf{k}-\frac{\mathbf{Q}}{2}}\right\}\right) \\
&-\frac{e}{2}\mathbf{B}\cdot\left(\mathbf{R_C}\times\sum_{\mathbf{Q}}\sum_{\mathbf{Q}'} w^*(\mathbf{Q}-\mathbf{Q_c})w(\mathbf{Q}'-\mathbf{Q_c})\delta_{\mathbf{Q},\mathbf{Q}'}\sum_{vc\mathbf{k}}\left\{\sum_{c'} A^{n\mathbf{Q}*}_{vc\mathbf{k}}A^{n\mathbf{Q}}_{vc'\mathbf{k}}\mathbf{v}_{cc'\mathbf{k}+\frac{\mathbf{Q}}{2}} - \sum_{v'} A^{n\mathbf{Q}*}_{vc\mathbf{k}}A^{n\mathbf{Q}}_{v'c\mathbf{k}}\mathbf{v}_{v'v\mathbf{k}-\frac{\mathbf{Q}}{2}}\right\}\right)
\end{aligned}
\quad \text{(S27)}
$$

From Eq. S23, it is clear that $\mathbf{R_C}$ has two contributions. The first contribution will cancel out the divergent term, involving $\nabla_{\mathbf{Q}}\delta_{\mathbf{Q},\mathbf{Q}'}$, in above expression. Subsequently, taking the limit $|w(\mathbf{Q}-\mathbf{Q_c})|^2 \rightarrow \delta(\mathbf{Q}-\mathbf{Q_c})$, we recover the final, gauge-invariant expression for the exciton orbital magnetic moment presented in Eq. 4.

## 5. Gauge invariance of exciton magnetic moment

We demonstrate here that the central expression for the exciton magnetic moment in Eq. 4 is invariant under arbitrary gauge transformations of the single-particle and exciton wavefunctions (3):

$$
\begin{aligned}
|u_{c\mathbf{k}}\rangle &\rightarrow e^{i\theta_c(\mathbf{k})}|u_{c\mathbf{k}}\rangle \\
|u_{v\mathbf{k}}\rangle &\rightarrow e^{i\theta_v(\mathbf{k})}|u_{v\mathbf{k}}\rangle \\
A^{n\mathbf{Q}}_{vc\mathbf{k}} &\rightarrow e^{i\theta_v\left(\mathbf{k}-\frac{\mathbf{Q}}{2}\right)}e^{-i\theta_c\left(\mathbf{k}+\frac{\mathbf{Q}}{2}\right)}e^{i\theta_n(\mathbf{Q})}A^{n\mathbf{Q}}_{vc\mathbf{k}} \\
|u^X_{n\mathbf{Q}}\rangle &\rightarrow e^{i\theta_n(\mathbf{Q})}|u^X_{n\mathbf{Q}}\rangle.
\end{aligned}
\quad \text{(S28)}
$$

We analyze the transformation properties of each term in the expression, focusing on the half of Eq. 4 involving conduction band indices (the terms involving valence bands follow analogously). Key identities under the gauge transformation include:

The exciton envelope function products and derivatives:

$$A^{n\mathbf{Q}*}_{vc\mathbf{k}}A^{n\mathbf{Q}}_{vc'\mathbf{k}} \rightarrow e^{i\theta_c\left(\mathbf{k}+\frac{\mathbf{Q}}{2}\right)}e^{-i\theta_{c'}\left(\mathbf{k}+\frac{\mathbf{Q}}{2}\right)}A^{n\mathbf{Q}*}_{vc\mathbf{k}}A^{n\mathbf{Q}}_{vc'\mathbf{k}},$$

$$A^{n\mathbf{Q}*}_{vc\mathbf{k}}\nabla_\mathbf{k}A^{n\mathbf{Q}}_{vc'\mathbf{k}} \rightarrow e^{i\theta_c\left(\mathbf{k}+\frac{\mathbf{Q}}{2}\right)}e^{-i\theta_{c'}\left(\mathbf{k}+\frac{\mathbf{Q}}{2}\right)}[i\nabla_\mathbf{k}\theta_v\left(\mathbf{k}-\frac{\mathbf{Q}}{2}\right)A^{n\mathbf{Q}*}_{vc\mathbf{k}}A^{n\mathbf{Q}}_{vc'\mathbf{k}} - i\nabla_\mathbf{k}\theta_{c'}\left(\mathbf{k}+\frac{\mathbf{Q}}{2}\right)A^{n\mathbf{Q}*}_{vc\mathbf{k}}A^{n\mathbf{Q}}_{vc'\mathbf{k}}$$
$$+A^{n\mathbf{Q}*}_{vc\mathbf{k}}\nabla_\mathbf{k}A^{n\mathbf{Q}}_{vc'\mathbf{k}}], \quad \text{(S29)}$$
$$A^{n\mathbf{Q}*}_{vc\mathbf{k}}\nabla_\mathbf{Q}A^{n\mathbf{Q}}_{vc'\mathbf{k}} \rightarrow e^{i\theta_c\left(\mathbf{k}+\frac{\mathbf{Q}}{2}\right)}e^{-i\theta_{c'}\left(\mathbf{k}+\frac{\mathbf{Q}}{2}\right)}[-\frac{i}{2}\nabla_\mathbf{k}\theta_v\left(\mathbf{k}-\frac{\mathbf{Q}}{2}\right)A^{n\mathbf{Q}*}_{vc\mathbf{k}}A^{n\mathbf{Q}}_{vc'\mathbf{k}} - \frac{i}{2}\nabla_\mathbf{k}\theta_{c'}\left(\mathbf{k}+\frac{\mathbf{Q}}{2}\right)A^{n\mathbf{Q}*}_{vc\mathbf{k}}A^{n\mathbf{Q}}_{vc'\mathbf{k}}$$
$$i\nabla_\mathbf{Q}\theta_n(\mathbf{Q})A^{n\mathbf{Q}*}_{vc\mathbf{k}}A^{n\mathbf{Q}}_{vc'\mathbf{k}} + A^{n\mathbf{Q}*}_{vc\mathbf{k}}\nabla_\mathbf{Q}A^{n\mathbf{Q}}_{vc'\mathbf{k}}].$$

The gauge transformations of the Berry connections and velocity matrix elements:

$$\boldsymbol{\mathcal{A}}_{cc'\mathbf{k}} = i\langle u_{c\mathbf{k}}|\nabla_\mathbf{k}u_{c'\mathbf{k}}\rangle \rightarrow e^{-i\theta_c(\mathbf{k})}e^{i\theta_{c'}(\mathbf{k})}[-\nabla_\mathbf{k}\theta_{c'}(\mathbf{k})\delta_{cc'} + \boldsymbol{\mathcal{A}}_{cc'\mathbf{k}}]$$
$$\boldsymbol{\mathcal{A}}^{\mathrm{X}}_{n\mathbf{Q}} = i\langle u^X_{n\mathbf{Q}}|\nabla_\mathbf{Q}u^X_{n\mathbf{Q}}\rangle \rightarrow -\nabla_\mathbf{Q}\theta_n(\mathbf{Q}) + i\langle u^X_{n\mathbf{Q}}|\nabla_\mathbf{Q}u^X_{n\mathbf{Q}}\rangle = -\nabla_\mathbf{Q}\theta_n(\mathbf{Q}) + \boldsymbol{\mathcal{A}}^{\mathrm{X}}_{n\mathbf{Q}}, \quad \text{(S30)}$$
$$\mathbf{v}_{cc'\mathbf{k}} = \frac{1}{\hbar}(\epsilon_{c'\mathbf{k}} - \epsilon_{c\mathbf{k}})\langle u_{c\mathbf{k}}|\nabla_\mathbf{k}u_{c'\mathbf{k}}\rangle \rightarrow \frac{1}{\hbar}(\epsilon_{c'\mathbf{k}} - \epsilon_{c\mathbf{k}})e^{-i\theta_c(\mathbf{k})}e^{i\theta_{c'}(\mathbf{k})}\langle u_{c\mathbf{k}}|\nabla_\mathbf{k}u_{c'\mathbf{k}}\rangle = e^{-i\theta_c(\mathbf{k})}e^{i\theta_{c'}(\mathbf{k})}\mathbf{v}_{cc'\mathbf{k}}.$$

The single-particle magnetic moment transformation:

$$\boldsymbol{\mu}_{cc'\mathbf{k}} = \frac{e}{2i\hbar}\langle\nabla_\mathbf{k}u_{c\mathbf{k}}| \times [H_\mathbf{k} - \epsilon_{c\mathbf{k}}]|\nabla_\mathbf{k}u_{c'\mathbf{k}}\rangle \rightarrow$$
$$\frac{e}{2i\hbar}e^{-i\theta_c(\mathbf{k})}e^{i\theta_{c'}(\mathbf{k})}[\nabla_\mathbf{k}\theta_c(\mathbf{k}) \times \nabla_\mathbf{k}\theta_{c'}(\mathbf{k})\langle u_{c\mathbf{k}}|(H_\mathbf{k} - \epsilon_{c\mathbf{k}})|u_{c'\mathbf{k}}\rangle$$
$$-i\nabla_\mathbf{k}\theta_c(\mathbf{k}) \times \langle u_{c\mathbf{k}}|(H_\mathbf{k} - \epsilon_{c\mathbf{k}})|\nabla_\mathbf{k}u_{c'\mathbf{k}}\rangle + i\langle\nabla_\mathbf{k}u_{c\mathbf{k}}|(H_\mathbf{k} - \epsilon_{c\mathbf{k}})|u_{c'k}\rangle \times \nabla_\mathbf{k}\theta_{c'}(\mathbf{k}) \quad \text{(S31)}$$
$$+\langle\nabla_\mathbf{k}u_{c\mathbf{k}}| \times [H_\mathbf{k} - \epsilon_{c\mathbf{k}}]|\nabla_\mathbf{k}u_{c'\mathbf{k}}\rangle].$$

Using relations $\langle u_{c\mathbf{k}}|(H_\mathbf{k} - \epsilon_{c\mathbf{k}}) = \langle u_{c\mathbf{k}}|(\epsilon_{c\mathbf{k}} - \epsilon_{c\mathbf{k}}) = 0$, and $(H_\mathbf{k} - \epsilon_{c\mathbf{k}})|u_{c'\mathbf{k}}\rangle = (\epsilon_{c'\mathbf{k}} - \epsilon_{c\mathbf{k}})|u_{c'\mathbf{k}}\rangle$,

$$\boldsymbol{\mu}_{cc'\mathbf{k}} \rightarrow e^{-i\theta_c(\mathbf{k})}e^{i\theta_{c'}(\mathbf{k})}\left[\frac{e}{2}\{\nabla_\mathbf{k}\theta_{c'}(\mathbf{k}) \times \mathbf{v}_{cc'\mathbf{k}}\} + \mathbf{m}_{cc'\mathbf{k}}\right]. \quad \text{(S32)}$$

Upon substituting these into the full expression and collecting terms, it can be confirmed that $\nabla_\mathbf{Q}\theta_n(\mathbf{Q})$ cancels between the terms involving $A^{n\mathbf{Q}*}_{vc\mathbf{k}}\nabla_\mathbf{Q}A^{n\mathbf{Q}}_{vc'\mathbf{k}}$ and $\boldsymbol{\mathcal{A}}^{\mathrm{X}}_{n\mathbf{Q}}$; $\nabla_\mathbf{k}\theta_v\left(\mathbf{k}-\frac{\mathbf{Q}}{2}\right)$ cancels between the terms involving $A^{n\mathbf{Q}*}_{vc\mathbf{k}}\nabla_\mathbf{k}A^{n\mathbf{Q}}_{vc'\mathbf{k}}$ and $A^{n\mathbf{Q}*}_{vc\mathbf{k}}\nabla_\mathbf{Q}A^{n\mathbf{Q}}_{vc'\mathbf{k}}$; while $\nabla_\mathbf{k}\theta_{c'}\left(\mathbf{k}+\frac{\mathbf{Q}}{2}\right)$ cancels among the terms involving $A^{n\mathbf{Q}*}_{vc\mathbf{k}}\nabla_\mathbf{k}A^{n\mathbf{Q}}_{vc'\mathbf{k}}$, $A^{n\mathbf{Q}*}_{vc\mathbf{k}}\nabla_\mathbf{Q}A^{n\mathbf{Q}}_{vc'\mathbf{k}}$, and $\boldsymbol{\mu}_{cc'\mathbf{k}}$ when $c \neq c'$; for $c = c'$, $\nabla_\mathbf{k}\theta_{c'}\left(\mathbf{k}+\frac{\mathbf{Q}}{2}\right)$ cancels among the terms involving $A^{n\mathbf{Q}*}_{vc\mathbf{k}}\nabla_\mathbf{k}A^{n\mathbf{Q}}_{vc'\mathbf{k}}$, $A^{n\mathbf{Q}*}_{vc\mathbf{k}}\nabla_\mathbf{Q}A^{n\mathbf{Q}}_{vc'\mathbf{k}}$, and $\boldsymbol{\mathcal{A}}_{cc'\mathbf{k}}$. Similar cancellation occurs for the other half of Eq. 4, making the whole expression gauge invariant. In order to compute the contribution of each term to the total exciton magnetic moment we choose a physical gauge under which the phase of the envelope function of *s*-exciton is uniform.

## 6. Approximations in previous literature

In the present formulation, the position operator in the Bloch basis is treated using its periodic analogue. This treatment is required because the exact position-operator matrix element contains both a singular intra-band contribution and a regular Berry-connection contribution, $\langle n'\mathbf{k}'|\hat{\mathbf{r}}|n\mathbf{k}\rangle = -i\delta_{nn'}(\nabla_{\mathbf{k}}\delta_{\mathbf{k}\mathbf{k}'}) + i\delta_{\mathbf{k}\mathbf{k}'}\langle u_{n'\mathbf{k}}|\nabla_{\mathbf{k}}u_{n\mathbf{k}}\rangle$ as derived in SM Sec. 4. The first term, proportional to $\nabla_{\mathbf{k}}\delta_{\mathbf{k}\mathbf{k}'}$, reflects the extended nature of Bloch states and requires special treatment. Introducing the periodic analogue of the position operator allows this contribution to be handled rigorously.

In Ref. (4), the position-operator matrix element is effectively restricted to its regular inter-band component,

$$\left\langle n'\mathbf{k}'|\hat{\mathbf{r}}|n\mathbf{k}\right\rangle \approx i\delta_{\mathbf{k}\mathbf{k}'}\langle u_{n'\mathbf{k}}|\nabla_{\mathbf{k}}u_{n\mathbf{k}}\rangle, \qquad n \neq n'. \tag{S33}$$

For nondegenerate bands, this inter-band Berry connection may be expressed in terms of the velocity matrix element as

$$i\langle u_{n'\mathbf{k}}|\nabla_{\mathbf{k}}u_{n\mathbf{k}}\rangle = i\hbar\frac{\langle n'\mathbf{k}|\hat{\mathbf{v}}|n\mathbf{k}\rangle}{\epsilon_{n\mathbf{k}} - \epsilon_{n'\mathbf{k}}}, \quad n \neq n' \tag{S34}$$

using $\langle n'\mathbf{k}|\hat{\mathbf{v}}|n\mathbf{k}\rangle = \frac{i}{\hbar}\langle n'\mathbf{k}|[H,\hat{\mathbf{r}}]|n\mathbf{k}\rangle$. These relations apply only to the inter-band sector and do not include the singular intra-band term. Within this approximation, inserting a complete single-particle basis at fixed $\mathbf{k}$: $\sum_m |m\mathbf{k}\rangle\langle m\mathbf{k}| = 1$, into Eq. 3 yields, for an exciton band $n$ at $\mathbf{Q} = 0$

$$\begin{aligned}\tilde{\epsilon}^X_{n\mathbf{Q=0}} &= \epsilon^X_{n\mathbf{Q=0}} \\ &+\frac{e}{2}\mathbf{B}\cdot\sum_{vc\mathbf{k}} A^*_{vc\mathbf{k}}[\sum_{c'\mathbf{k}'m} A_{vc'\mathbf{k}'}\langle c\mathbf{k}|\hat{\mathbf{r}}|m\mathbf{k}\rangle \times \langle m\mathbf{k}|\hat{\mathbf{v}}|c'\mathbf{k}'\rangle - \sum_{v'\mathbf{k}'m} A_{v'c\mathbf{k}'}\langle v'\mathbf{k}'|\hat{\mathbf{r}}|m\mathbf{k}\rangle \times \langle m\mathbf{k}|\hat{\mathbf{v}}|v\mathbf{k}\rangle] \\ &\sim\epsilon^X_{n\mathbf{Q=0}} + \frac{e}{2}i\hbar\mathbf{B}\cdot\sum_{vc\mathbf{k}} A^*_{vc\mathbf{k}}[\sum_{c',m\neq c} A_{vc'\mathbf{k}}\frac{\langle c\mathbf{k}|\hat{\mathbf{v}}|m\mathbf{k}\rangle \times \langle m\mathbf{k}|\hat{\mathbf{v}}|c'\mathbf{k}\rangle}{\epsilon_{m\mathbf{k}} - \epsilon_{c\mathbf{k}}} - \sum_{v',m\neq v'} A_{v'c\mathbf{k}}\frac{\langle v'\mathbf{k}|\hat{\mathbf{v}}|m\mathbf{k}\rangle \times \langle m\mathbf{k}|\hat{\mathbf{v}}|v\mathbf{k}\rangle}{\epsilon_{m\mathbf{k}} - \epsilon_{v'\mathbf{k}}}].\end{aligned} \tag{S35}$$

This is the gauge-invariant multiband expression employed in Ref. (4). Its derivation, however, does not retain the singular intra-band part of the position operator. Such intra-band contributions are known to play an essential role in the semiclassical dynamics and quantum geometric response of Bloch electrons (5).

A different approximation is adopted in Ref. (6), where the exciton energy shift is written as an ansatz to be

$$\tilde{\epsilon}^X_{n\mathbf{Q}} = \epsilon^X_{n\mathbf{Q}} + \frac{e}{2}\mathbf{B}\cdot\sum_{vc\mathbf{k}}\left|A^{n\mathbf{Q}}_{vc\mathbf{k}}\right|^2[\langle c\mathbf{k}+\mathbf{Q}|\hat{\mathbf{r}}\times\hat{\mathbf{v}}|c\mathbf{k}+\mathbf{Q}\rangle - \langle v\mathbf{k}|\hat{\mathbf{r}}\times\hat{\mathbf{v}}|v\mathbf{k}\rangle]\,. \tag{S36}$$

This form amounts to retaining only the part of the orbital-moment operator that is diagonal in both crystal momentum and band index, $\langle n'\mathbf{k}'|\hat{\mathbf{r}}\times\hat{\mathbf{v}}|n\mathbf{k}\rangle \sim \delta_{\mathbf{k}\mathbf{k}'}\delta_{nn'}\langle n\mathbf{k}|\hat{\mathbf{r}}\times\hat{\mathbf{v}}|n\mathbf{k}\rangle = \delta_{\mathbf{k}\mathbf{k}'}\delta_{nn'}\frac{1}{i\hbar}\langle\nabla_{\mathbf{k}}u_{i\mathbf{k}}|\times[H_{\mathbf{k}} - \epsilon_{i\mathbf{k}}]|\nabla_{\mathbf{k}}u_{j\mathbf{k}}\rangle$. It therefore omits off-diagonal band-coherence contributions to the exciton magnetic response.

In contrast, our present derivation retains fully both the singular intra-band and regular inter-band parts of the position operator. Their consistent treatment generates the Berry-phase-corrected single-particle, relative-motion, and COM quantum-geometric contributions, whose sum gives the fully gauge-invariant exciton orbital magnetic moment (Eq. 4).

## 7. Methods

Ab initio calculations of the electronic structure of BBG were performed using density functional theory (DFT) within the local density approximation (LDA) for the exchange-correlation functional, as implemented in the *Quantum ESPRESSO* package (*7*). Optimized norm-conserving Vanderbilt pseudopotentials were employed (*8*). The Kohn-Sham orbitals were expanded in a plane-wave basis with a kinetic energy cutoff of 70 Ry. To eliminate spurious interactions between periodic images, we applied a truncated Coulomb interaction in the out-of-plane direction (*9*) and included a vacuum spacing of 20 Å, following extensive convergence tests. Structural relaxation for pristine BBG was carried out until the forces on atoms were below $10^{-4}\ eV/\text{Å}$, yielding an in-plane lattice constant of $a = 2.448$ Å (see Fig. S2).

To simulate hBN encapsulation, monolayers of hBN were placed above and below BBG, assuming a commensurate in-plane lattice constant. The out-of-plane atomic positions were then relaxed to the same force tolerance as above, yielding interlayer distances of 3.329 Å between carbon layers and 3.216 Å between carbon and hBN layers. Among the various high-symmetry stacking configurations, the most energetically favorable arrangement places boron directly over the carbon sites, while nitrogen avoids direct overlap with carbon atoms in AB-stacked BBG. Depending on whether boron lies above/below A or B sublattices of graphene, there are four possible configurations, one of which is used in Fig. 2a, where boron atoms are placed over the non-dimer sites of BBG. It should be noted that in pristine BBG the valence and conduction band edges are contributed by the non-dimer sites (*10*). In Fig. S3, we show results for an alternative stacking in which the boron atom sits over a dimer site on one side and a non-dimer site on the other.

While DFT-LDA provides reliable Kohn-Sham orbitals as quasiparticle wavefunctions, it underestimates band gaps and band dispersion. To address this issue, we carried out single-shot $G_0W_0$ calculations (*1*) using the *BerkeleyGW* package (*11*) to obtain quasiparticle self-energy corrections to the band energies. The static dielectric matrix was computed within the random phase approximation using a plane-wave cutoff of 8 Ry and 1,000 unoccupied bands. To accelerate convergence with respect to BZ sampling, we employed the non-uniform neck subsampling (NNS) technique (*12*). A $24 \times 24 \times 1$ Monkhorst-Pack grid with ten NNS points was used for dielectric function and self-energy calculations. Due to the huge band gap of hBN, the effect of hBN on the dielectric screening of BG is small, hence we assume the dielectric matrix to be the same for pristine BBG and BBG with hBN. The resulting quasiparticle band structure (Fig. 2b) was interpolated using maximally localized Wannier functions generated with the *Wannier90* code (*13*).

Excitonic properties were computed by solving the BSE on top of the GW-corrected band structure (*14*), again using the *BerkeleyGW* package (*11*). We used a patched sampling scheme focusing on the K and K′ valleys, achieving an effective resolution equivalent to a $900 \times 900$ uniform mesh, which ensured convergence of exciton energies to within 5 meV. The e-h interaction kernel was evaluated using a dielectric matrix computed with a plane-wave cutoff of 6 Ry and 350 unoccupied bands. For finite COM momentum **Q** excitons (*15*), we performed BSE calculations using two high-resolution patches centered around K and K′ valleys, enabling accurate inclusion of inter-valley exchange scattering. The BSE Hamiltonian is constructed using three valence and three conduction bands. The contribution of each of the valence-conduction pairs to the 1s exciton wavefunction is

| $\sum_{\mathbf{k}} \lvert A_{v_n c_n \mathbf{k}} \rvert^2$ | $c_1$ | $c_2$ | $c_3$ |
|---|---|---|---|
| $v_1$ | 0.978 | 0.009 | 0.000 |
| $v_2$ | 0.013 | 0.000 | 0.000 |
| $v_3$ | 0.000 | 0.000 | 0.000 |

where $v_1$ denotes the highest occupied valence band and $c_1$ the lowest unoccupied conduction band. The dominant contribution (~98%) from the $v_1c_1$ pair reflects the fact that these transitions have the smallest electron-hole excitation energies (Fig. S1), while contributions from all other inter-band pairs are negligible. This demonstrates the convergence of our results with respect to the number of inter-band pairs included in the calculation, while also suggesting that the two-band approximation for the exciton orbital magnetic moment calculation might provide a decent description for this particular system.

Note that we first construct and solve the BSE Hamiltonian at finite exciton COM momentum **Q** in the absence of a magnetic field to obtain the unperturbed exciton energies and wavefunctions at different **Q** in the Brillouin zone. The exciton orbital magnetic moment is then evaluated with perturbation theory using the expressions derived in this work. This perturbative treatment, together with the fixed-screening approximation, allows the first-order magnetic response to be expressed entirely in terms of the **Q**-dependent unperturbed exciton energies and wavefunctions.

To assess the effect of hBN encapsulation and stacking, we compared excitonic orbital magnetic moments for pristine BBG and both hBN/BBG configurations as described above. hBN encapsulation introduces sublattice symmetry breaking in each graphene layer, enhancing particle–hole asymmetry and increasing the difference between conduction and valence band single-particle orbital magnetic moments. This trend is evident in the comparison between Fig. S2 (pristine) and Fig. 2c (encapsulated), leading to a significant impact on the valley g-factors. Further we compare values for the two hBN/BBG configurations studied – i.e., values in Table 1 vs values in Fig. S3. For the two configurations, the *s*-exciton g-factors are 13.48 and 20.32, while the *p*-exciton values are 2.58 and 1.76. In both cases, the disparity between *s*- and *p*-exciton magnetic responses persists, and the newly derived contributions dominate the final value of the g-factor for the case of *p*-exciton. Among the two stacking configurations, the one with boron on the non-dimer carbon sites shows better agreement with experiment. While our simulations do not account for the incommensurability between graphene and hBN lattices, this comparison suggests that such a stacking arrangement may have been prominent in experimental samples.

## 8. Single-particle magnetic moment

Instead of calculating the magnetic moment directly for a free quasiparticle in band $n$ and wavevector $\mathbf{k}$ using the form given by

$$[\boldsymbol{\mu}_{n\mathbf{k}}]_z = \frac{e}{2i\hbar}[\langle\nabla_{\mathbf{k}}u_{n\mathbf{k}}|\times(H_{\mathbf{k}}-\epsilon_{n\mathbf{k}})|\nabla_{\mathbf{k}}u_{n\mathbf{k}}\rangle]_z, \tag{S37}$$

where $H_{\mathbf{k}}$ is the GW quasiparticle Hamiltonian, we recast it by inserting a complete basis set, yielding the more computationally convenient form

$$[\boldsymbol{\mu}_{n\mathbf{k}}]_z = \frac{e}{2i\hbar}\sum_j\left[\epsilon_{j\mathbf{k}}-\epsilon_{n\mathbf{k}}\right]\left[\boldsymbol{\mathcal{A}}^*_{jn\mathbf{k}}\times\boldsymbol{\mathcal{A}}_{jn\mathbf{k}}\right]_z, \tag{S38}$$

where $\epsilon_{n\mathbf{k}}$ is the quasiparticle energy, and $\boldsymbol{\mathcal{A}}_{jn\mathbf{k}} = i\langle u_{j\mathbf{k}}|\nabla_{\mathbf{k}}u_{n\mathbf{k}}\rangle$ is the single-particle band Berry connection. We find that for BBG, it is essential to include remote bands outside the low-energy four-band subspace in the summation over bands $j$ (see Fig. S1), indicating that effective four-band models commonly used in the literature (*16*) may be insufficient to capture accurately the single-particle orbital magnetic response of this material.

## 9. Exciton orbital moment for a general exciton wavefunction definition

In Eq. 2, we defined the exciton wavefunction using the symmetric momentum-partition convention, in which the conduction- and valence-band crystal momenta in the expression are shifted by $+\mathbf{Q}/2$ and $-\mathbf{Q}/2$, respectively. More generally, without changing any physics (it is a matter of relabeling of basis functions), the exciton COM momentum $\mathbf{Q}$ may be distributed arbitrarily between the electron and hole single-particle states. For example, one may assign the entire $\mathbf{Q}$ to the electron, $\mathbf{k_c} = \mathbf{k}+\mathbf{Q}$ and $\mathbf{k_v} = \mathbf{k}$, or to the hole, $\mathbf{k_c} = \mathbf{k}$ and $\mathbf{k_v} = \mathbf{k}-\mathbf{Q}$. A general momentum-partition convention can be written as

$$\begin{aligned} |X_{n\mathbf{Q}}\rangle &= \sum\nolimits_{vc\mathbf{k}} A^{n\mathbf{Q}}_{vc\mathbf{k}}|c,\mathbf{k}+(1-\alpha)\mathbf{Q}\rangle\otimes|v,\mathbf{k}-\alpha\mathbf{Q}\rangle^*, \\ \langle\mathbf{r_e},\mathbf{r_h}|X_{n\mathbf{Q}}\rangle &= e^{i\mathbf{Q}\cdot\left(\frac{\mathbf{r_e}+\mathbf{r_h}}{2}\right)}\sum\nolimits_{vc\mathbf{k}} A^{n\mathbf{Q}}_{vc\mathbf{k}}e^{i\left(\mathbf{k}+\left(\frac{1}{2}-\alpha\right)\mathbf{Q}\right)\cdot(\mathbf{r_e}-\mathbf{r_h})}u_{c,\mathbf{k}+(1-\alpha)\mathbf{Q}}(\mathbf{r_e})u^*_{v,\mathbf{k}-\alpha\mathbf{Q}}(\mathbf{r_h}), \end{aligned} \tag{S39}$$

where $0 \le \alpha \le 1$. The symmetric convention used in Eq. 2 corresponds to $\alpha = 1/2$. The momentum conjugate to the relative coordinate $\mathbf{r_e}-\mathbf{r_h}$ is then $\mathbf{k_{rel}} = \frac{\mathbf{k_v}+\mathbf{k_c}}{2} = \mathbf{k}+\left(\frac{1}{2}-\alpha\right)\mathbf{Q}$. Changing $\alpha$ therefore corresponds to the reparameterization of the relative-momentum coordinate rather than a physical modification of the exciton state. Using Eq. S39, we first evaluate the perturbed exciton Hamiltonian in the basis of the unperturbed exciton eigenstates, following the procedure described in SM Sec. 2. This gives

$$
\begin{aligned}
&\langle X_{n\mathbf{Q}}|\tilde{H}^X|X_{n'\mathbf{Q}'}\rangle = \delta_{\mathbf{Q},\mathbf{Q}'}\delta_{n,n'}\epsilon^X_{n\mathbf{Q}} - \\
&\mathbf{B}\cdot\sum_{vc\mathbf{k}}[\sum_{c'\mathbf{k}'} A^{n\mathbf{Q}*}_{vc\mathbf{k}} A^{n'\mathbf{Q}'}_{vc'\mathbf{k}'}\delta_{\mathbf{k}-\alpha\mathbf{Q},\mathbf{k}'-\alpha\mathbf{Q}'}\langle c,\mathbf{k}+(1-\alpha)\mathbf{Q}|\hat{\boldsymbol{\mu}}|c',\mathbf{k}'+(1-\alpha)\mathbf{Q}'\rangle \\
&-\sum_{v'\mathbf{k}'} A^{n\mathbf{Q}*}_{vc\mathbf{k}} A^{n'\mathbf{Q}'}_{v'c\mathbf{k}'}\delta_{\mathbf{k}+(1-\alpha)\mathbf{Q},\mathbf{k}'+(1-\alpha)\mathbf{Q}'}\langle v',\mathbf{k}'-\alpha\mathbf{Q}'|\hat{\boldsymbol{\mu}}|v,\mathbf{k}-\alpha\mathbf{Q}\rangle].
\end{aligned} \tag{S40}
$$

The momentum-conservation conditions appearing in Eq. S40 now depend explicitly on $\alpha$. Next, we replace $\hat{\mathbf{r}}$-operator in $\hat{\boldsymbol{\mu}} = -\frac{e}{2}\hat{\mathbf{r}}\times\hat{\mathbf{v}}$ with its periodic analogue $\hat{\mathbf{r}}_{\boldsymbol{\lambda}}{}' = -i\nabla_{\boldsymbol{\lambda}} e^{i\boldsymbol{\lambda}\cdot\hat{\mathbf{r}}}$ to obtain

$$
\begin{aligned}
&\langle X_{n\mathbf{Q}}|\tilde{H}^X|X_{n'\mathbf{Q}'}\rangle = \delta_{\mathbf{Q},\mathbf{Q}'}\epsilon^X_{n\mathbf{Q}} + \\
&\lim_{\boldsymbol{\lambda}\to 0}\frac{e}{2i}\mathbf{B}\cdot(\nabla_{\boldsymbol{\lambda}}\times\sum_{vc\mathbf{k}}[\sum_{c'\mathbf{k}'} A^{n\mathbf{Q}*}_{vc\mathbf{k}} A^{n'\mathbf{Q}'}_{vc'\mathbf{k}'}\delta_{\mathbf{k}-\alpha\mathbf{Q},\mathbf{k}'-\alpha\mathbf{Q}'}\langle c,\mathbf{k}+(1-\alpha)\mathbf{Q}|e^{i\boldsymbol{\lambda}\cdot\hat{\mathbf{r}}}\hat{\mathbf{v}}|c',\mathbf{k}'+(1-\alpha)\mathbf{Q}'\rangle \\
&-\sum_{v'\mathbf{k}'} A^{n\mathbf{Q}*}_{vc\mathbf{k}} A^{n'\mathbf{Q}'}_{v'c\mathbf{k}'}\delta_{\mathbf{k}+(1-\alpha)\mathbf{Q},\mathbf{k}'+(1-\alpha)\mathbf{Q}'}\langle v',\mathbf{k}'-\alpha\mathbf{Q}'|e^{i\boldsymbol{\lambda}\cdot\hat{\mathbf{r}}}\hat{\mathbf{v}}|v,\mathbf{k}-\alpha\mathbf{Q}\rangle])
\end{aligned} \tag{S41}
$$

Using the matrix element from SM Sec. 3, $\langle n\mathbf{k}|e^{i\boldsymbol{\lambda}\cdot\hat{\mathbf{r}}}\hat{\mathbf{v}}|n'\mathbf{k}'\rangle = \delta_{\mathbf{k}',\mathbf{k}-\boldsymbol{\lambda}}\langle n\mathbf{k}|\hat{\mathbf{v}}(\mathbf{k}-\boldsymbol{\lambda})|n'\mathbf{k}-\boldsymbol{\lambda}\rangle$ with $\hat{\mathbf{v}}(\mathbf{k}) = e^{-i\mathbf{k}\cdot\mathbf{r}}\hat{\mathbf{v}}e^{i\mathbf{k}\cdot\mathbf{r}}$, produces $\alpha$-dependent relations among $\mathbf{Q}$, $\mathbf{Q}'$, $\mathbf{k}$, and $\mathbf{k}'$. Differentiation with respect to $\boldsymbol{\lambda}$ followed by limit $\boldsymbol{\lambda}\to 0$ generates corresponding $\alpha$-dependent coefficients. The divergence associated with the exciton COM coordinate is then treated using the exciton wave-packet construction described in SM Sec. 4.

The center of the wave packet remains that given in Eq. S23, while the exciton-band Berry connection takes the general form

$$
\boldsymbol{\mathcal{A}}^{\mathrm{X}}_{n\mathbf{Q}} = i\sum_{vc\mathbf{k}} A^{n\mathbf{Q}*}_{vc\mathbf{k}}\nabla_{\mathbf{Q}} A^{n\mathbf{Q}}_{vc\mathbf{k}} + \sum_{vck}\left\{(1-\alpha)\sum_{c'} A^{n\mathbf{Q}*}_{vc\mathbf{k}} A^{n\mathbf{Q}}_{vc'\mathbf{k}}\boldsymbol{\mathcal{A}}_{cc',\mathbf{k}+(1-\alpha)\mathbf{Q}} + \alpha\sum_{v'} A^{n\mathbf{Q}*}_{vc\mathbf{k}} A^{n\mathbf{Q}}_{v'ck}\boldsymbol{\mathcal{A}}_{v'v,\mathbf{k}-\alpha\mathbf{Q}}\right\}. \tag{S42}
$$

Following the same procedure as in the symmetric derivation, we obtain the exciton orbital magnetic moment for a general momentum-partition convention:

$$
\begin{aligned}
\boldsymbol{\mu}^X_{n\mathbf{Q}} &= \sum_{vc\mathbf{k}}\left\{\sum_{c'} A^{n\mathbf{Q}*}_{vc\mathbf{k}} A^{n\mathbf{Q}}_{vc'\mathbf{k}}\boldsymbol{\mu}_{cc',\mathbf{k}+(1-\alpha)\mathbf{Q}} - \sum_{v'} A^{n\mathbf{Q}*}_{vc\mathbf{k}} A^{n\mathbf{Q}}_{v'c\mathbf{k}}\boldsymbol{\mu}'_{v'v,\mathbf{k}-\alpha\mathbf{Q}}\right\} \\
&+\frac{e}{2\hbar}\sum_{vc\mathbf{k}}\left\{\sum_{c'} A^{n\mathbf{Q}*}_{vc\mathbf{k}} A^{n\mathbf{Q}}_{vc'\mathbf{k}}(\nabla_{\mathbf{k}}\epsilon_{c\mathbf{k}}\times\boldsymbol{\mathcal{A}}_{cc'\mathbf{k}})|_{\mathbf{k}+(1-\alpha)\mathbf{Q}} - \sum_{v'} A^{n\mathbf{Q}*}_{vc\mathbf{k}} A^{n\mathbf{Q}}_{v'c\mathbf{k}}(\nabla_{\mathbf{k}}\epsilon_{v\mathbf{k}}\times\boldsymbol{\mathcal{A}}_{v'v\mathbf{k}})|_{\mathbf{k}-\alpha\mathbf{Q}}\right\} \\
&+\frac{e}{2i}\sum_{vc\mathbf{k}}\left\{\sum_{c'}\alpha\left(A^{n\mathbf{Q}*}_{vc\mathbf{k}}\nabla_{\mathbf{k}} A^{n\mathbf{Q}}_{vc'\mathbf{k}}\times\mathbf{v}_{cc',\mathbf{k}+(1-\alpha)\mathbf{Q}}\right) + \sum_{v'}(1-\alpha)\left(A^{n\mathbf{Q}*}_{vc\mathbf{k}}\nabla_{\mathbf{k}} A^{n\mathbf{Q}}_{v'c\mathbf{k}}\times\mathbf{v}_{v'v,\mathbf{k}-\alpha\mathbf{Q}}\right)\right\} \\
&+\frac{e}{2i}\sum_{vc\mathbf{k}}\left\{\sum_{c'}\left(A^{n\mathbf{Q}*}_{vc\mathbf{k}}\mathbf{D}_{\mathbf{Q}} A^{n\mathbf{Q}}_{vc'\mathbf{k}}\times\mathbf{v}_{cc',\mathbf{k}+(1-\alpha)\mathbf{Q}}\right) - \sum_{v'}\left(A^{n\mathbf{Q}*}_{vc\mathbf{k}}\mathbf{D}_{\mathbf{Q}} A^{n\mathbf{Q}}_{v'c\mathbf{k}}\times\mathbf{v}_{v'v,\mathbf{k}-\alpha\mathbf{Q}}\right)\right\}
\end{aligned} \tag{S43}
$$

where $\mathbf{D_Q}$ is the covariant derivative with respect to exciton COM momentum given by $\mathbf{D_Q} = \boldsymbol{\nabla}_\mathbf{Q} + i\boldsymbol{\mathcal{A}}^X_{n\mathbf{Q}}$. Eq. S43 is invariant under arbitrary gauge transformations of the single-particle Bloch states and the exciton state,

$$\begin{aligned} |u_{c\mathbf{k}}\rangle &\to e^{i\theta_c(\mathbf{k})}|u_{c\mathbf{k}}\rangle \\ |u_{v\mathbf{k}}\rangle &\to e^{i\theta_v(\mathbf{k})}|u_{v\mathbf{k}}\rangle \\ A^{n\mathbf{Q}}_{vc\mathbf{k}} &\to e^{i\theta_v(\mathbf{k}-\alpha\mathbf{Q})}e^{-i\theta_c(\mathbf{k}+(1-\alpha)\mathbf{Q})}e^{i\theta_n(\mathbf{Q})}A^{n\mathbf{Q}}_{vc\mathbf{k}} \\ |u^X_{n\mathbf{Q}}\rangle &\to e^{i\theta_n(\mathbf{Q})}|u^X_{n\mathbf{Q}}\rangle. \end{aligned} \tag{S44}$$

The transformation of $A^{n\mathbf{Q}}_{vc\mathbf{k}}$ under single-particle gauge rotations is explicitly $\alpha$-dependent. Nevertheless, the cancellation of all arbitrary gauge phases proceeds in the same manner as described in SM Sec. 5. Although Eq. S43 contains explicit $\alpha$-dependent factors, this dependence arises solely from the chosen parametrization of the relative momentum. To show this explicitly, we introduce the momentum variable $\bar{\mathbf{k}} = \mathbf{k} + \left(\frac{1}{2} - \alpha\right)\mathbf{Q}$. The exciton amplitudes and their derivatives then transform as

$$\begin{aligned} \boldsymbol{\nabla}_\mathbf{k} A^{n\mathbf{Q}}_{vc\mathbf{k}} &= \boldsymbol{\nabla}_{\bar{\mathbf{k}}} A^{n\mathbf{Q}}_{vc\bar{\mathbf{k}}} \\ \boldsymbol{\nabla}_\mathbf{Q} A^{n\mathbf{Q}}_{vc\mathbf{k}} &= \boldsymbol{\nabla}_\mathbf{Q} A^{n\mathbf{Q}}_{vc\bar{\mathbf{k}}} + \left(\frac{1}{2} - \alpha\right)\boldsymbol{\nabla}_{\bar{\mathbf{k}}} A^{n\mathbf{Q}}_{vc\bar{\mathbf{k}}} \end{aligned} \tag{S45}$$

Substituting Eq. S45 into Eq. S43 and changing the summation variable from $\mathbf{k}$ to $\bar{\mathbf{k}}$ gives:

$$\begin{aligned} \boldsymbol{\mu}^X_{n\mathbf{Q}} = &\sum_{vc\bar{\mathbf{k}}}\left\{\sum_{c'} A^{n\mathbf{Q}*}_{vc\bar{\mathbf{k}}}A^{n\mathbf{Q}}_{vc'\bar{\mathbf{k}}}\boldsymbol{\mu}_{cc',\bar{\mathbf{k}}+\frac{\mathbf{Q}}{2}} - \sum_{v'} A^{n\mathbf{Q}*}_{vc\bar{\mathbf{k}}}A^{n\mathbf{Q}}_{v'c\bar{\mathbf{k}}}\boldsymbol{\mu}'_{v'v,\bar{\mathbf{k}}-\frac{\mathbf{Q}}{2}}\right\} \\ &+\frac{e}{2\hbar}\sum_{vc\bar{\mathbf{k}}}\left\{\sum_{c'} A^{n\mathbf{Q}*}_{vc\bar{\mathbf{k}}}A^{n\mathbf{Q}}_{vc'\bar{\mathbf{k}}}(\boldsymbol{\nabla}_{\bar{\mathbf{k}}}\epsilon_{c\bar{\mathbf{k}}} \times \boldsymbol{\mathcal{A}}_{cc'\bar{\mathbf{k}}})|_{\bar{\mathbf{k}}+\frac{\mathbf{Q}}{2}} - \sum_{v'} A^{n\mathbf{Q}*}_{vc\bar{\mathbf{k}}}A^{n\mathbf{Q}}_{v'c\bar{\mathbf{k}}}(\boldsymbol{\nabla}_{\bar{\mathbf{k}}}\epsilon_{v\bar{\mathbf{k}}} \times \boldsymbol{\mathcal{A}}_{v'v\bar{\mathbf{k}}})|_{\bar{\mathbf{k}}-\frac{\mathbf{Q}}{2}}\right\} \\ &+\frac{e}{4i}\sum_{vc\bar{\mathbf{k}}}\left\{\sum_{c'}\left(A^{n\mathbf{Q}*}_{vc\bar{\mathbf{k}}}\boldsymbol{\nabla}_{\bar{\mathbf{k}}}A^{n\mathbf{Q}}_{vc'\bar{\mathbf{k}}} \times \mathbf{v}_{cc',\bar{\mathbf{k}}+\frac{\mathbf{Q}}{2}}\right) + \sum_{v'}\left(A^{n\mathbf{Q}*}_{vc\bar{\mathbf{k}}}\boldsymbol{\nabla}_{\bar{\mathbf{k}}}A^{n\mathbf{Q}}_{v'c\bar{\mathbf{k}}} \times \mathbf{v}_{v'v,\bar{\mathbf{k}}-\frac{\mathbf{Q}}{2}}\right)\right\} \\ &+\frac{e}{2i}\sum_{vc\bar{\mathbf{k}}}\left\{\sum_{c'}\left(A^{n\mathbf{Q}*}_{vc\bar{\mathbf{k}}}\mathbf{D_Q}A^{n\mathbf{Q}}_{vc'\bar{\mathbf{k}}} \times \mathbf{v}_{cc',\bar{\mathbf{k}}+\frac{\mathbf{Q}}{2}}\right) - \sum_{v'}\left(A^{n\mathbf{Q}*}_{vc\bar{\mathbf{k}}}\mathbf{D_Q}A^{n\mathbf{Q}}_{v'c\bar{\mathbf{k}}} \times \mathbf{v}_{v'v,\bar{\mathbf{k}}-\frac{\mathbf{Q}}{2}}\right)\right\} \end{aligned} \tag{S46}$$

Eq. S46 is identical to the result obtained using the symmetric convention. In particular, the additional term generated by the transformation of $\boldsymbol{\nabla}_\mathbf{Q} A^{n\mathbf{Q}}_{vc\mathbf{k}}$ exactly combines with the explicitly $\alpha$-dependent relative-motion terms in Eq. S43, recovering the symmetric coefficient $e/4i$. This explicitly demonstrates that the total exciton orbital magnetic moment is independent of the momentum-partition parameter $\alpha$. The symmetric convention is therefore a matter of convenience rather than a physical assumption.

## 10. <u>Sensitivity of results on $G_0W_0$ starting point</u>

The weak dependence of the computed values of the exciton magnetic moment in BBG on the DFT mean-field starting point for the $G_0W_0$ quasiparticle calculation can be understood from its low-energy electronic structure. The bands relevant to the excitons considered here belong to the carbon-$p_z$ manifold (Fig. S1a). Their sublattice and layer hybridization is primarily governed by the lattice geometry, interlayer coupling, and applied bias. Changing the exchange-correlation approximation therefore only slightly renormalizes the DFT parameters within the same low-energy manifold rather than changing its orbital character or qualitatively reconstructing the corresponding wavefunctions. Since at the $G_0W_0$ level the mean-field DFT energies and wavefunctions are used as approximations to the quasiparticle quantities in the construction of the self-energy operator, we consequently expect the quantities entering the exciton magnetic moment to be relatively insensitive to the exchange-correlation functional used.

We explicitly verify this by calculating the DFT single-particle orbital magnetic moments of the relevant valence and conduction bands using LDA, PBE, and SCAN. As shown in Fig. S5, the difference between the conduction and valence-band orbital moments, which directly enters the single-particle contribution to the exciton magnetic moment, exhibits very little variation among the three functionals. The individual band moments likewise show only weak dependence on the exchange-correlation approximation.

We have also performed $G_0W_0$ calculations starting from each of the three functionals. As shown in Fig. S6, the resulting quasiparticle band dispersions remain very similar, with the quasiparticle band gap varying by only a few meV among the three starting points. Thus, both the wavefunction-dependent quantum geometric quantities and the quasiparticle energies and band dispersion relevant to our full exciton magnetic-moment calculation are only very weakly affected by the choice among these starting points.

Finally, the exciton envelope functions are obtained from the Bethe-Salpeter equation, whose matrix elements depend on the underlying single-particle wavefunctions, screened electron-hole interaction, and quasiparticle dispersions. Given the very similar low-energy wavefunctions and quasiparticle band structures obtained from the three starting points, we expect the resulting exciton wavefunctions, and hence the calculated exciton valley g-factors, to exhibit correspondingly weak dependence on the choice of starting point.

## 11. Supplementary figures

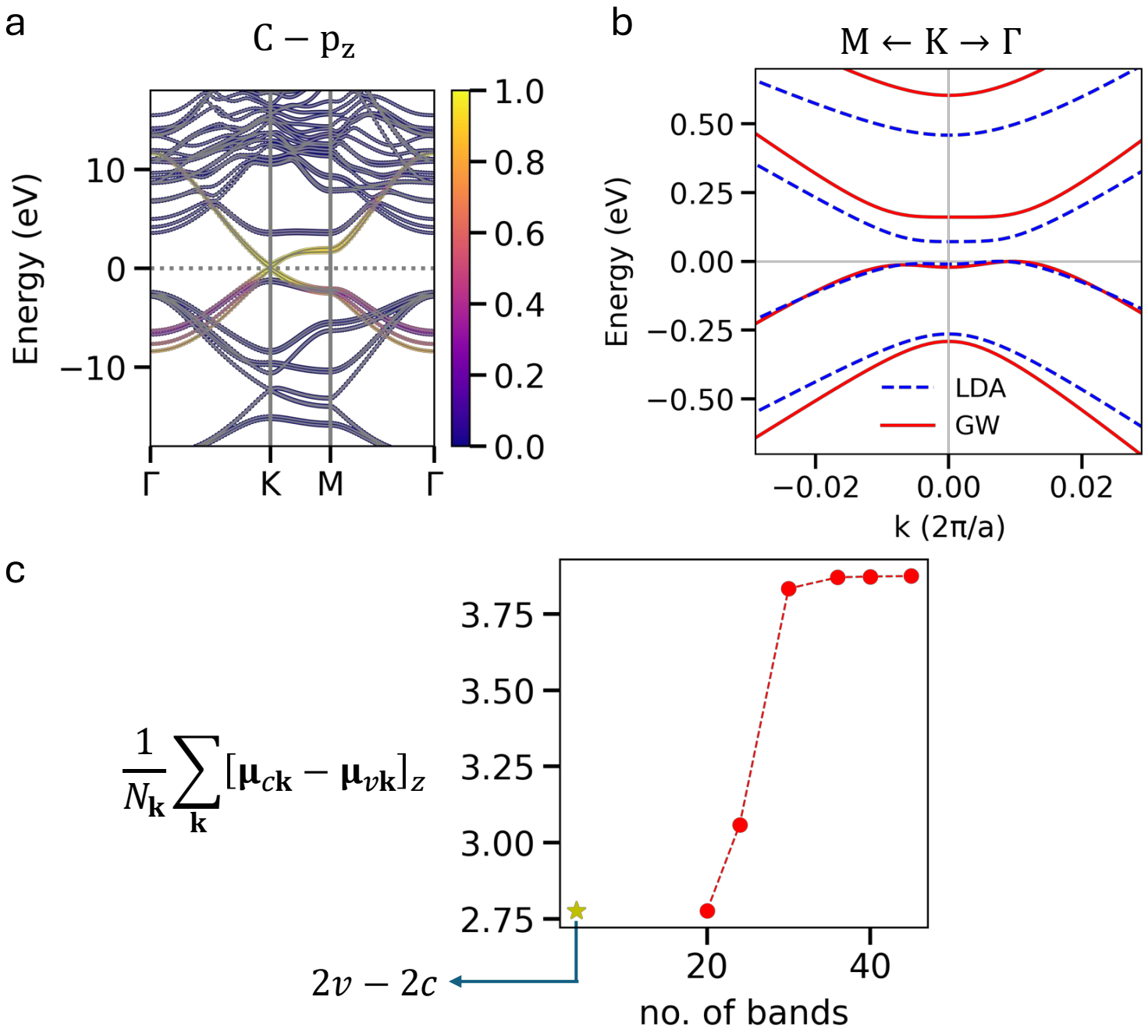

**Fig. S1: Single-particle band structure and convergence of the single-particle magnetic moment. a**, First-principles electronic band structure of BBG (with bias field of 0.10 eV/Å ) with hBN encapsulation calculated using DFT-LDA along high-symmetry lines in the Brillouin zone. The color scale indicates contributions from the $p_z$ orbitals of the four carbon atoms in the unit cell. The low-energy states near the Fermi level arise entirely from these orbitals, motivating the use of an effective four-band model in many studies in the literature. **b**, Zoomed-in band structure near the K point comparing LDA (dashed blue) with GW (solid red) results. **c**, Convergence of the difference in single-particle orbital magnetic moments between the conduction and valence bands as a function of the number of bands included in the summation in Eq. S38. The yellow star indicates the result obtained using only the four bands in conventional effective models, which significantly underestimates the fully converged value, underscoring the importance of the remote bands away from the fundamental band gap.

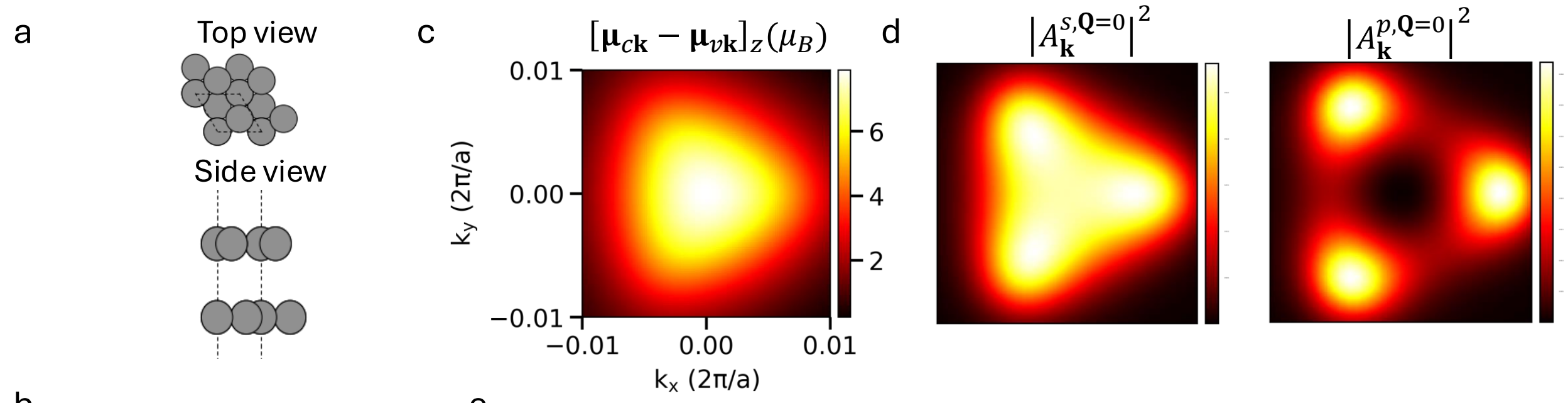

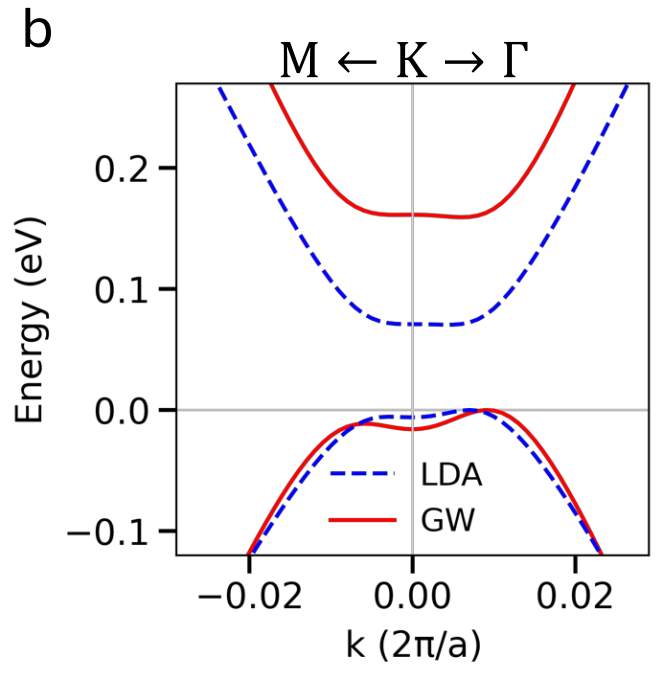

e

| | s | p |
|---|---|---|
| Envelope function average | 5.20 | 4.13 |
| Berry phase correction to e and h moments | 1.33 | 2.06 |
| e-h relative moment | −0.06 | −3.53 |
| e-h COM moment | −0.68 | −2.59 |
| Total exciton orbital magnetic moment | 5.79 | 0.07 |
| Exciton valley g-factor | **11.58** | **0.14** |

**Fig. S2: Exciton magnetic moment contributions for pristine BBG with bias field of $\mathbf{0.10\ eV/Å}$. a**, Atomic configuration of pristine AB-stacked BBG, that is, no hBN encapsulation. Grey color denotes carbon atoms. Dashed lines indicate the unit-cell. **b**, Electronic band structure of pristine BBG near the K-point calculated using DFT-LDA (blue dashed) and GW (red solid). **c**, Map of the difference between the conduction and valence band single-particle orbital magnetic moments (along $z$ direction) near the K-point in units of Bohr magneton ($\mu_B$). The (0,0) point corresponds to $\mathbf{k} = \mathrm{K}$. The values are significantly lower than those calculated for BBG with hBN encapsulation (Fig. 2c). **d**, Absolute amplitude square of the exciton **k**-space envelope functions near the K point for the $s$-exciton (left) and $p$-excitons (right). Axis labels are the same of those in **c**. **e**, Calculated exciton valley $g$-factor using Eq. 11 of main text (last row in the table) and contributions of various terms (first four rows) to the exciton orbital magnetic moment for the $s$ and $p$ exciton states at $\mathbf{Q} = 0$ corresponding to exciton localized in K valley. The values for the other valley excitons are negative of those shown here.

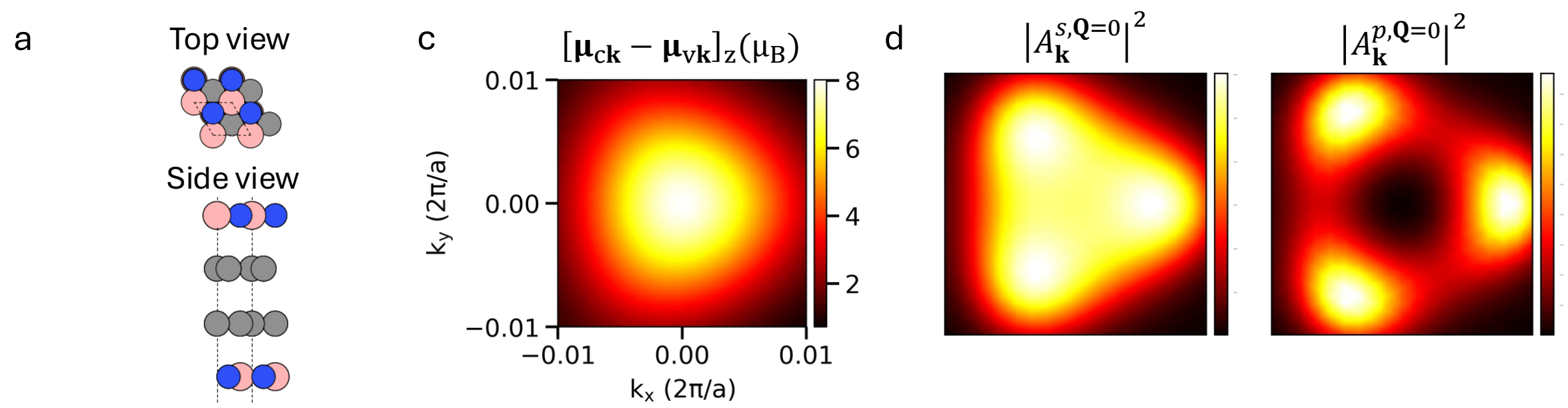

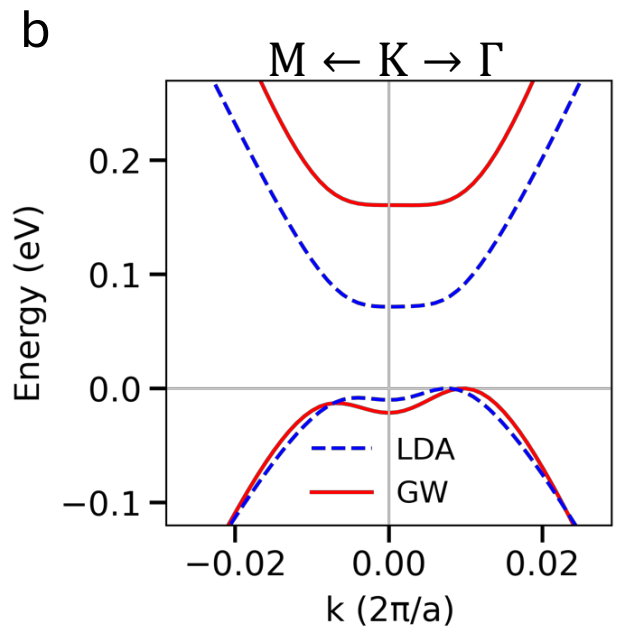

e

| | s | p |
|---|---|---|
| Envelope function average | 5.76 | 4.80 |
| Berry phase correction to e and h moments | 1.69 | 2.46 |
| e-h relative moment | −0.04 | −3.46 |
| e-h COM moment | −0.67 | −2.51 |
| Total exciton orbital magnetic moment | 6.74 | 1.29 |
| Exciton valley g-factor | **13.48** | **2.58** |

**Fig. S3: Exciton magnetic moment contributions for a BBG/hBN configuration with bias field of 0.10 eV/Å. a**, Atomic configuration of BBG with hBN monolayer on both sides. Grey color denotes carbon atoms, while blue and pink denote nitrogen and boron atoms respectively. Dashed lines indicate the unit-cell. Carbon atoms forming the dimer sites are placed on the dashed line, while the other carbon atom in each layer is on a non-dimer site. Boron is placed below the non-dimer site of BBG on one side while over the dimer site on the other. **b**, Electronic band structure from DFT-LDA (blue dashed) and from GW (red solid) calculated near the K-point. **c**, Momentum-resolved difference between the conduction and valence band single-particle orbital magnetic moments (along $z$) near the K-point in units of Bohr magneton ($\mu_B$). The (0,0) point corresponds to $\mathbf{k}$ = K. The values are higher than those calculated for pristine BBG (Fig. S2c). **d**, Momentum-resolved absolute amplitude square of the exciton envelope functions near the K-point for the $s$-exciton (left) and $p$-exciton (right). Axis labels match that of **c**. **e**, Calculated exciton valley $g$-factor using Eq. 11 of main text (last row in the table) and contributions of various terms (first four rows) to the exciton orbital magnetic moment for the $s$ and $p$ exciton states at $\mathbf{Q} = 0$ corresponding to exciton localized in K valley. The values for the other valley exciton are negative of those shown here.

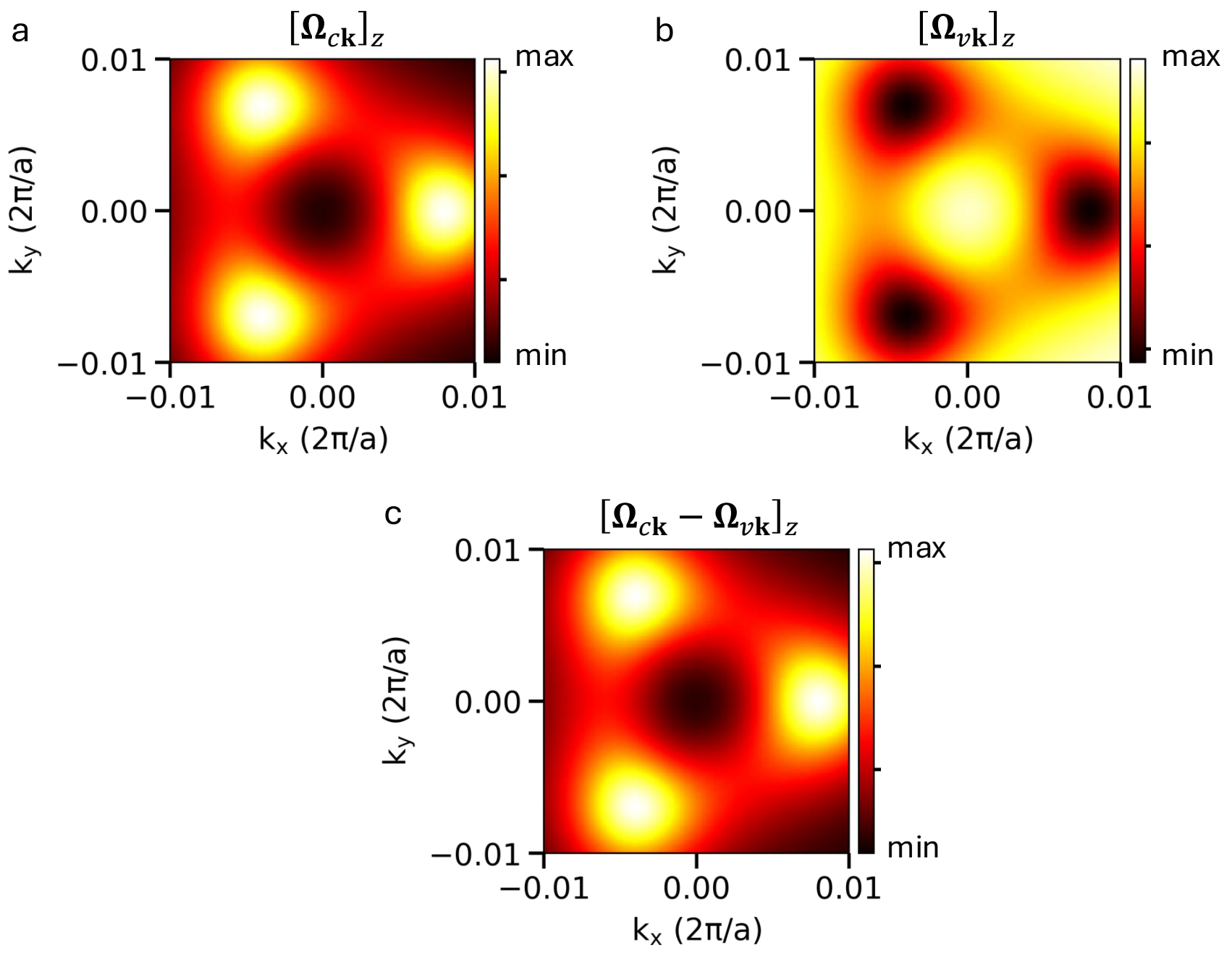

**Fig. S4: Berry curvatures of single-particle valence and conduction bands in BBG. a**, and **b**, The z component (out-of-plane) of the Berry curvature $\boldsymbol{\Omega}_{n\mathbf{k}} = \boldsymbol{\nabla}_{\boldsymbol{k}} \times \boldsymbol{\mathcal{A}}_{n\mathbf{k}}$ computed for the valence and conduction band of BBG with bias field of $0.10\ \mathrm{eV/Å}$, respectively. **c**, Difference between the z component of the Berry curvatures of the conduction and valence band. The resulting profile exhibits a distribution with peaked structures similar to those of the amplitude square of the *p*-exciton envelope function (Fig. 3c), indicating a larger Berry-phase correction to the magnetic moment of *p*-exciton compared to *s*-exciton in BBG. The origin of the plots, $\mathbf{k} = (0,0)$, is set at the $\mathbf{k} = \mathrm{K}$ point of the Brillouin zone.

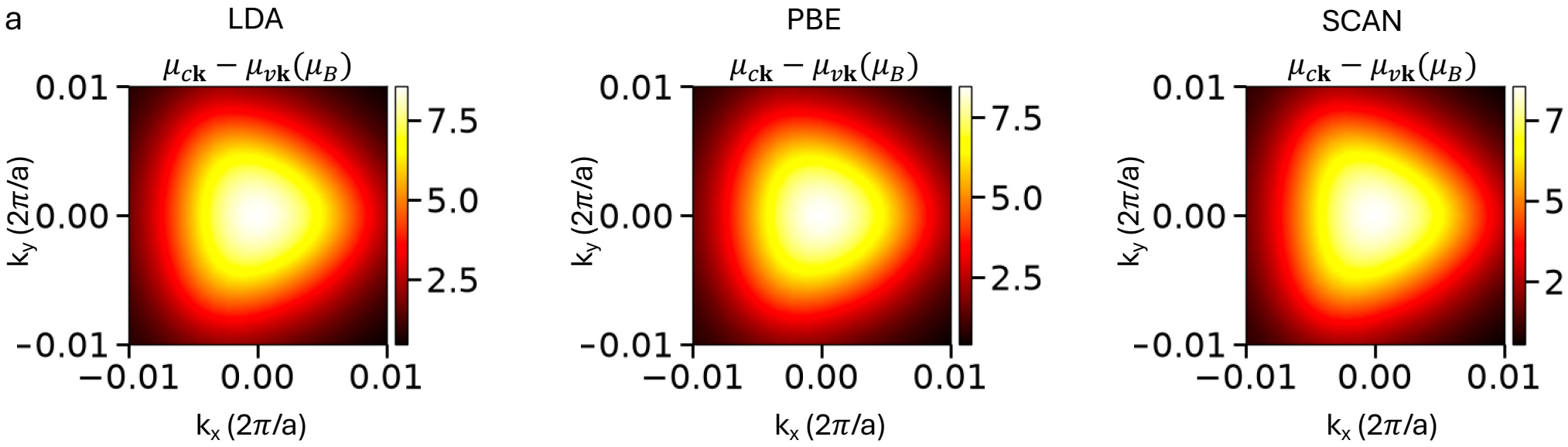

b

| Orbital moments ($\mu_B$) | LDA | PBE | SCAN |
|---|---|---|---|
| $\mu_{c\mathrm{K}}$ | -1.232 | -1.549 | -1.372 |
| $\mu_{v\mathrm{K}}$ | -9.793 | -10.027 | -9.921 |
| $\mu_{c\mathrm{K}} - \mu_{v\mathrm{K}}$ | 8.561 | 8.478 | 8.549 |

**Fig. S5. Comparison of the DFT single-particle orbital magnetic moment in BBG obtained using LDA, PBE, and SCAN mean-field starting points.** (a) Momentum-resolved orbital-

moment difference near K. (b) Orbital magnetic moments at K, showing weak dependence on the choice of exchange-correlation functional.

a

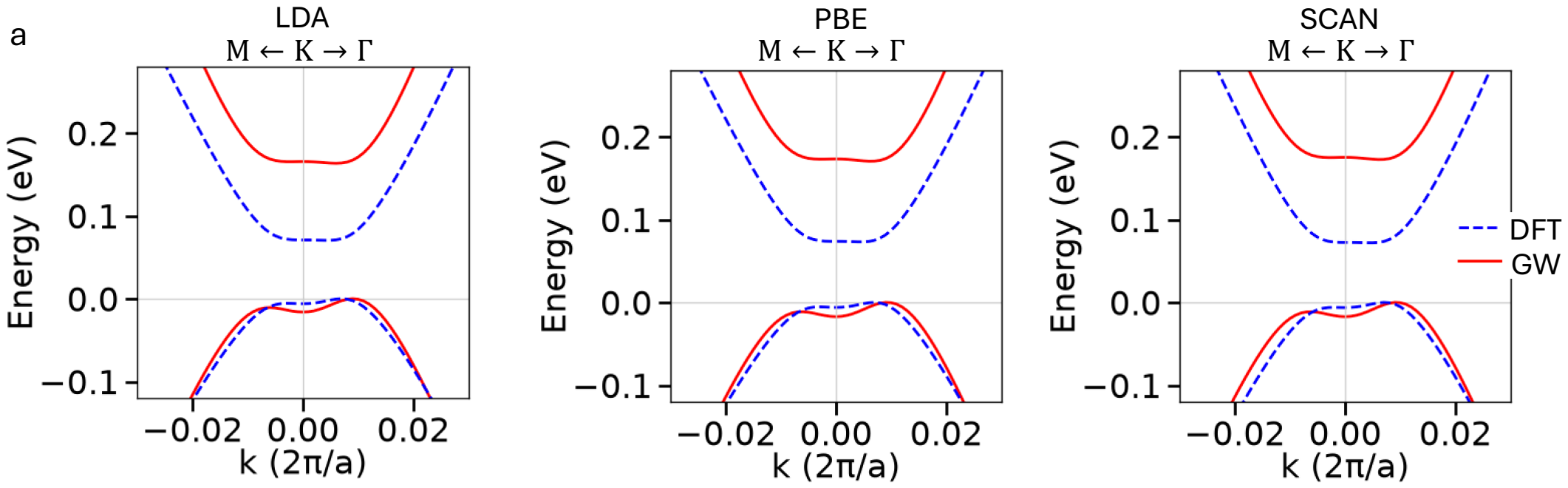

b

| | LDA | PBE | SCAN |
|---|---|---|---|
| DFT gap (meV) | 70.5 | 73.2 | 72.6 |
| $G_0W_0$ gap (meV) | 163 | 170 | 168 |

**Fig. S6. Comparison of the DFT and GW electronic structure of BBG obtained using LDA, PBE, and SCAN starting points.** (a) DFT and $G_0W_0$ band structures near K. (b) Corresponding DFT and $G_0W_0$ band gaps, showing weak starting-point dependence.